\documentclass[twocolumn,tradiabstract]{aa} 

\usepackage{natbib}
\usepackage{graphicx}
\usepackage{amsmath}
\usepackage[varg]{txfonts}

\newcommand\mancha{\textsc{Mancha3D~}}

\usepackage{txfonts}
\usepackage{xcolor}
\usepackage{soul}  
\setstcolor{red}

\usepackage[normalem]{ulem}

\begin{document}

\title{Convergence study of ambipolar diffusion in realistic simulations of magneto-convection}
\titlerunning{High-resolution non-ideal convection}

\author{E. Khomenko\inst{1,2} \and 
N. Vitas\inst{1,2} \and M. Collados\inst{1,2} \and M. Modestov\inst{1,2}}

\authorrunning{Khomenko et al.}

\institute{Instituto de Astrof\'{\i}sica de Canarias, 38205 La Laguna, Tenerife, Spain
\and Departamento de Astrof\'{\i}sica, Universidad de La Laguna, 38205, La Laguna, Tenerife, Spain}

\date{Received; Accepted }

\abstract {The aim of this paper is to improve our understanding of the heating mechanisms of the solar chromosphere via realistic three-dimensional (3D) modeling of solar magneto-convection, considering the fact that solar plasma contains a significant fraction of neutral gas. For that we performed simulations of the same physically volume of the Sun, namely $5.76\times 5.76\times2.3$ Mm$^3$ (with 1.4 Mm being above the optical surface), at three different resolutions: $20\times20\times14$, $10\times10\times7$ and $5\times5\times3.5$ km$^3$. At all three resolutions we compare the time series of simulations with/without ambipolar diffusion, as the main non-ideal heating  mechanism due to neutrals. We also compare simulations with three different magnetizations: (1) case of a small scale dynamo; (2) an initially implanted vertical magnetic field of 50 G; (3) an initially implanted vertical field of 200 G, though not all of them are available at all resolutions. We obtain that the average magnetization of the simulations increases with improving resolution, both in the dynamo and in the unipolar cases. So does the average magnetic Poynting flux, meaning that there is more magnetic energy in the simulation box at higher resolutions. Ambipolar diffusion operates at relatively large scales, which can be actually numerically resolved with the grid scale of the highest resolution simulations as the ones reported here. We consider two ways of evaluating where the ambipolar scales are numerically resolved. On the one hand, we provide a method to evaluate the numerical diffusion of the simulations and compare it to the physical ambipolar diffusion. On the other hand, we compare an order of magnitude evaluation of spatial scales given by the ambipolar diffusion to our grid resolution. This allows us to isolate locations where the  physical ambipolar effect is numerically resolved. At those locations we compare the average temperature in the simulations with/without ambipolar diffusion, and conclude that the plasma is on average about about 600 K hotter after 1200 sec of simulation time when the ambipolar diffusion is included. The amount of temperature enhancement increases with the resolution and with time, with no signs of saturation at our best horizontal resolution of 5 km.}


\maketitle

\section{Introduction} 

The question of chromospheric heating in solar physics remains a significant challenge. It is firmly established that the propagation and release of energy across the solar atmosphere are influenced by the substantial presence of neutral atoms in both the photosphere and the chromosphere \citep{Ballester+etal2018}. In a strongly collision-coupled solar plasma, various primary non-ideal effects, either originating from or influenced by neutrals, come into play: ambipolar diffusion, the Hall effect, and the Biermann battery effect. Significantly, it has been found that ambipolar diffusion facilitates energy transfer between magnetic and thermal energy by orders of magnitude. It is now proposed as a viable candidate to explain chromospheric heating \citep{Khomenko+Collados2012, Shelyag+etal2016, MartinezSykora+etal2012, Martinez-Sykora+etal2017, Martinez-Sykora+etal2020, Martinez-Sykora+etal2023, Nobrega-Siverio+etal2020, Khomenko+etal2018, Khomenko+etal2021, Gonzalez-Morales+etal2020}. The Biermann battery effect has been found useful for seeding magnetic fields for the small-scale dynamo, explaining the magnetization of the quiet Sun at a level suggested by observations \citep{Khomenko+etal2017}. The non-ideal Hall effect has been shown to generate Alfvén waves throughout the lower atmosphere, contributing to the chromospheric energy balance \citep{Cally+Khomenko2015, Gonzalez-Morales+etal2019, Gonzalez-Morales+etal2020}.

In our previous studies, we presented the first 3D realistic models of solar small-scale dynamo and magnetoconvection that incorporated non-ideal effects \citep{Khomenko+etal2018, Khomenko+etal2021, Gonzalez-Morales+etal2020}. The magnetic fields in the quiet Sun are most accurately represented by the small-scale dynamo model \citep{Rempel2014}. Conversely, regions like plage and magnetic network areas can be explained by a model where an initially unipolar magnetic field is introduced and evolves due to convection. Our simulations illustrated the basic principles of how the chromosphere of the quiet Sun can be heated. Surprisingly, we observed that weaker, more intricate small-scale dynamo fields have a more substantial impact on chromospheric heating through the ambipolar mechanism than vertically implanted fields. Following a thorough analysis, we attributed this phenomenon to the fact that energy in dynamo models is deposited deeper in the photosphere, directly contributing to a temperature increase. This is in contrast to the model with an implanted field, where a portion of the energy is used for ionization at the fronts of slow shock waves \citep{Khomenko+etal2018}. Additionally, our findings revealed that the Hall effect promotes the creation of long-lived vertical structures and facilitates the propagation of Alfv\'en waves into the chromosphere along these structures \citep{Gonzalez-Morales+etal2020}. This results in a consistent supply of magnetic energy to the chromosphere, which is then dissipated through the ambipolar effect, leading to a rise in temperature. These outcomes underscore the important role of quiet Sun magnetism in maintaining the chromospheric energy balance.

The realistic magneto-convection experiments mentioned above were made at a resolution of $20\times20\times14$ km$^3$. Theoretical calculations \citep{Khomenko+etal2014}, and the results by other authors \citep{Soler2019, Arber2016, Gonzalez-Morales+etal2019} suggest that the action of ambipolar diffusion and Hall effect increases with resolution due to the creation of small-scale currents. No convergence experiments for the action of ambipolar diffusion in realistic simulations of magneto-convection have been reported yet. \citet{Khomenko+Cally2019} performed convergence experiments in a setup more ideal for wave propagation in a scenario with an ensemble of small-scale flux tubes, showing that the ambipolar dissipation increases linearly with increasing resolution and that the results were not yet converged at 10 km resolution. The aim of the present work is to perform a convergence study of the heating rate in simulations of small-scale dynamo and magneto-convection by studying how the heating depends on the numerical resolution. Unlike the Ohmic diffusion, which operates at extremely short scales related to ion-electron collisional time scales \citep[below a fraction of a meter in the chromosphere, see][]{Khomenko+etal2014}, the ambipolar diffusion operates at scales related to ion-neutral collisions, which are significantly larger (reaching up to a kilometer). By increasing the spatial resolution of our simulations we hope to start numerically resolving the scales of the ambipolar diffusion in a larger volume of our simulation box. 

In our previous works \citep{Khomenko+etal2018, Khomenko+etal2021, Gonzalez-Morales+etal2020}, we computed long time series of simulations, with durations of 1-2 hours, where we switched on/off the ambipolar and the Hall effects. These models with/without a certain effect were run for sufficiently long time before the snapshots for the analysis were saved with a cadence of 10 sec. This way we aimed at having enough statistics to study the net effect on the magnetic Poynting flux absorption and heating. While this strategy worked well for a coarse horizontal numerical resolution of 20 km, the computing time and, most importantly, the size of the snapshots at a finer resolution, prevent this kind of analysis from being effectively done in practice. 
Another drawback of our previous analysis comes from the proper nature of solar convection, which is a stochastic non-linear process. When simulations with non-ideal effects are initiated from a given initial condition, this causes a perturbation, and a particular granulation structure diverges from the one of the run without non-ideal effects after about 10--15 min of solar time (granulation turnover time). In our previous experiments the time series could be compared only statistically, and a one to one comparison of the snapshots was not possible because of these changes of the granulation pattern. Here we aim at directly comparing snapshots with/without certain ingredients and at three different resolutions. For that we use shorter time series of simulations of each type starting from exactly the same initial conditions, to ensure the similarity between the structures that develop in these models. 

The paper is organized as follows. After describing the technical details about the simulations in Section \ref{sect:setup} and their average properties in Section \ref{sect:averages}, we proceed by discussing the methods to evaluate the artificial diffusion effects in the simulations and compare the numerical scales to those of the physical ambipolar diffusion in Section \ref{sect:heating}. Section \ref{sect:efficiency} shows the results of the computation of the heating rates, and Section \ref{sect:scaling} discusses the scaling with the resolution. The conclusions are given in Section \ref{sect:conclusions}.

\begin{table} 
\begin{center}
\caption{Duration/resolution of the simulation runs} 
\begin{tabular}{cccc} 
\hline Res/ID &  Dyn $\langle B_z\rangle$=0 G & $\langle B_z\rangle$=50 G & $\langle B_z\rangle$=200 G \\ \hline
20$\times$20$\times$14 km$^3$ &  1200 sec & 1580 sec &  1040 sec \\
 10$\times$10$\times$7 km$^3$ & 1200 sec & 870 sec &  $-$ \\
5$\times$5$\times$3.5  km$^3$ & 240 sec & 7 sec & $-$ \\
\hline
\end{tabular}
\end{center}
\label{tab:setup}
\end{table}

\begin{figure}
\centering
\includegraphics[width=9cm]{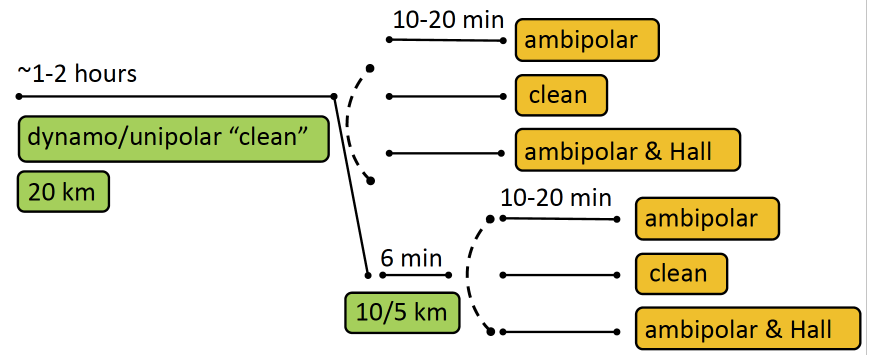}
\caption{\footnotesize Simulation setup. Green labels indicate time spans where the simulations were run to either reach the stationary state or settle the refined resolution. The yellow labels indicate the time spans where the snapshots were saved for the analysis every 10 sec.}
\label{fig:setup}
\end{figure}

\begin{figure*}
\centering
\includegraphics[keepaspectratio,width=6cm]{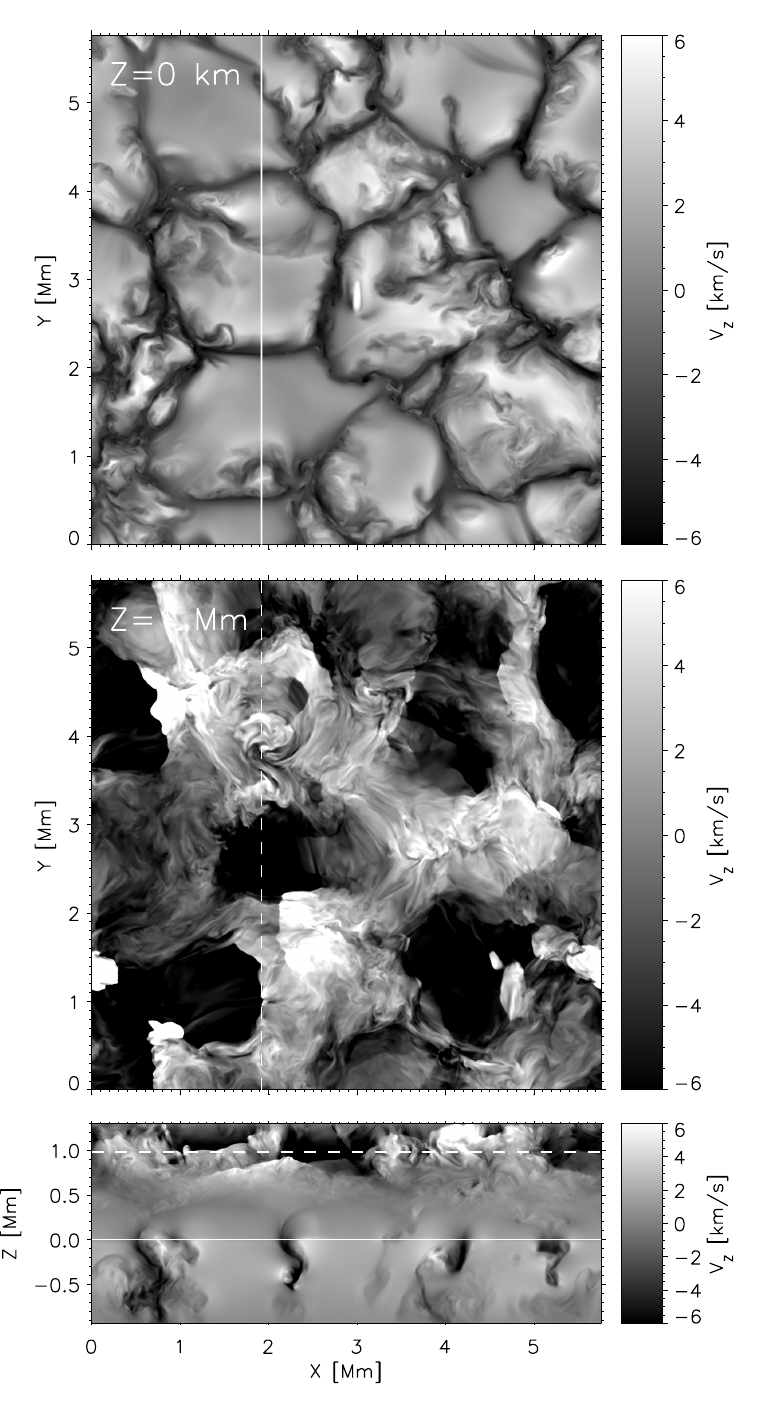}
\includegraphics[keepaspectratio,width=6cm]{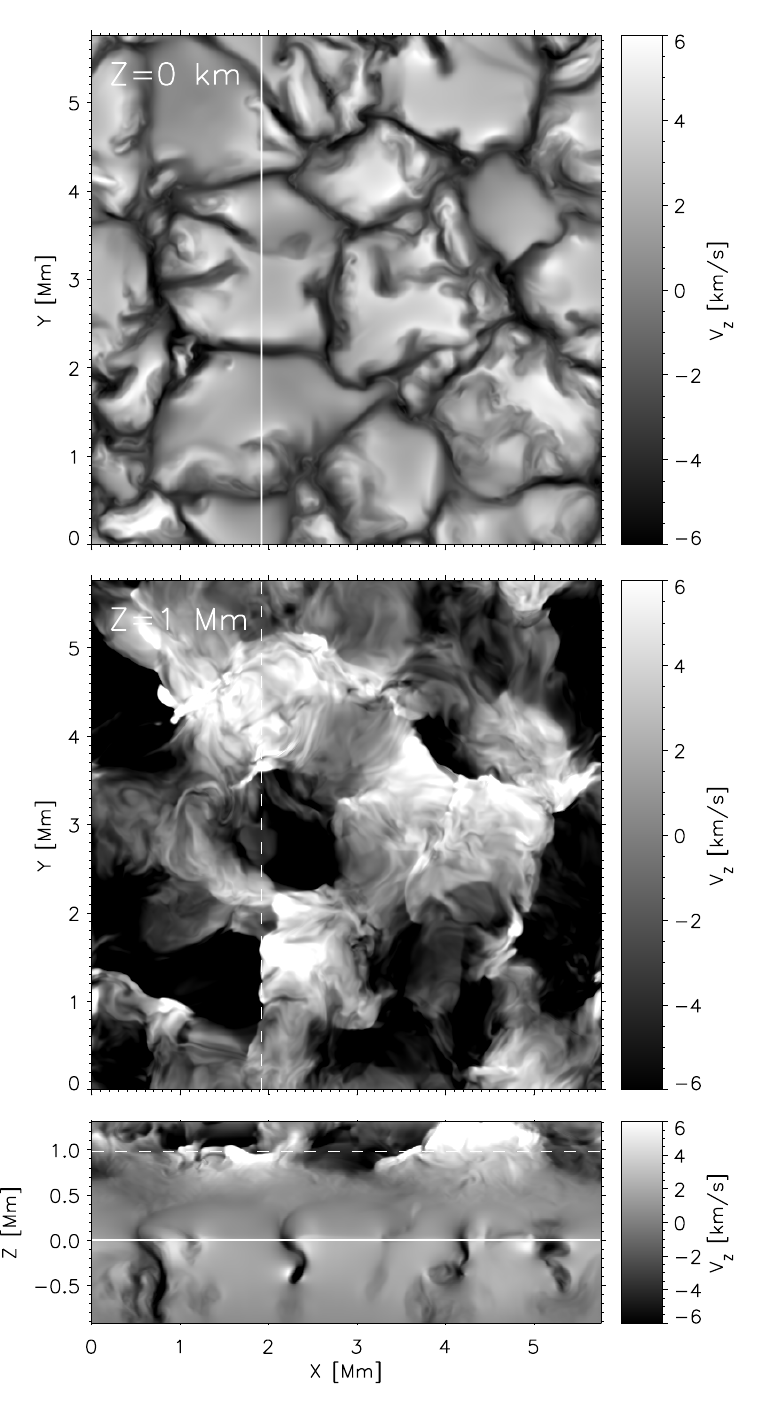}
\includegraphics[keepaspectratio,width=6cm]{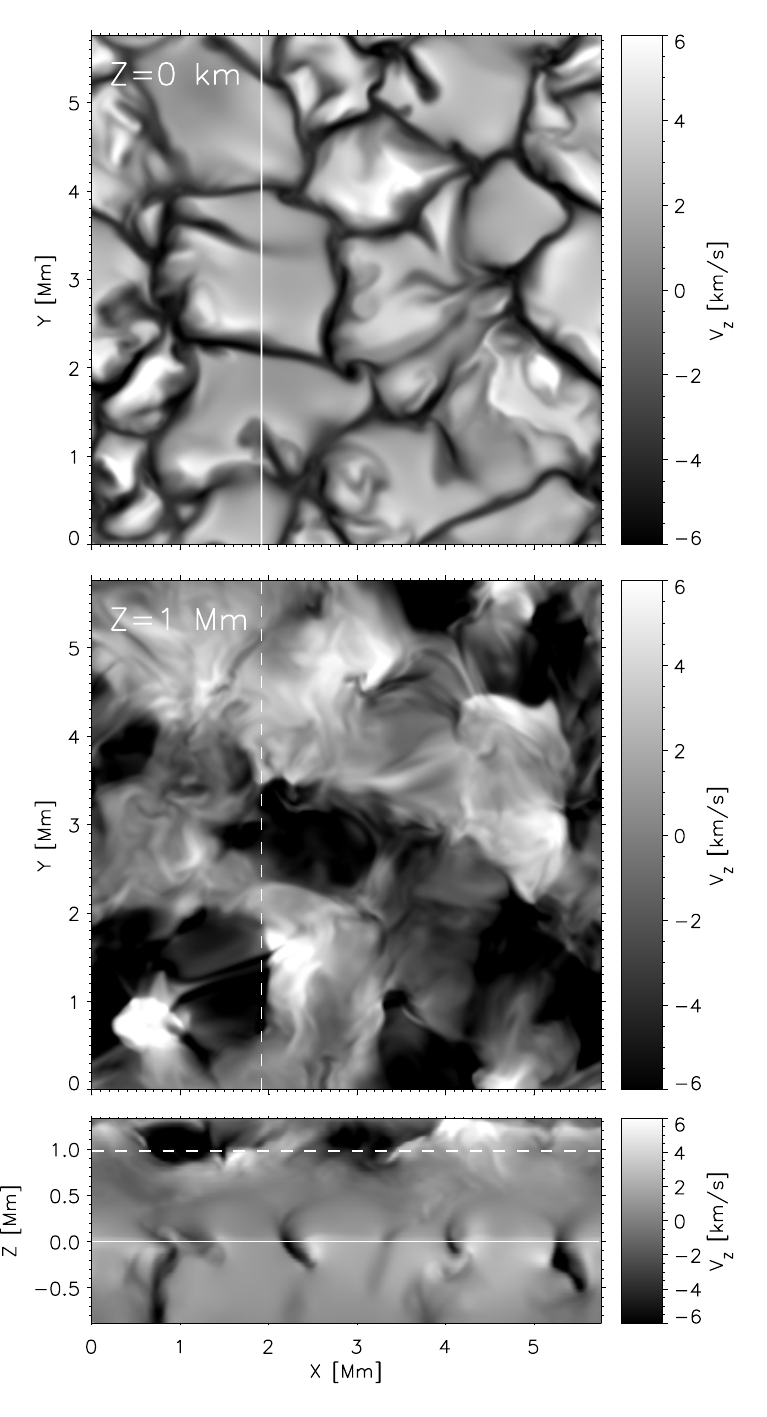}
\caption{\footnotesize Snapshots of the vertical velocities in the photosphere at $z=0$ km (top row), in the chromosphere at $z=1$ Mm (middle row) and in a vertical cut at an arbitrary horizontal location (bottom row). Horizontal lines at the top and middle panels mark the location of the slice from the bottom panel. Horizontal lines at the bottom panels mark the heights where the velocities at the upper panels are shown. Panels from left to right are for Dyn$_5$, Dyn$_{10}$ and Dyn$_{20}$ simulations, correspondingly. }
\label{fig:vzsnap-dyn}
\vspace{6truemm}
\centering
\includegraphics[keepaspectratio,width=6cm]{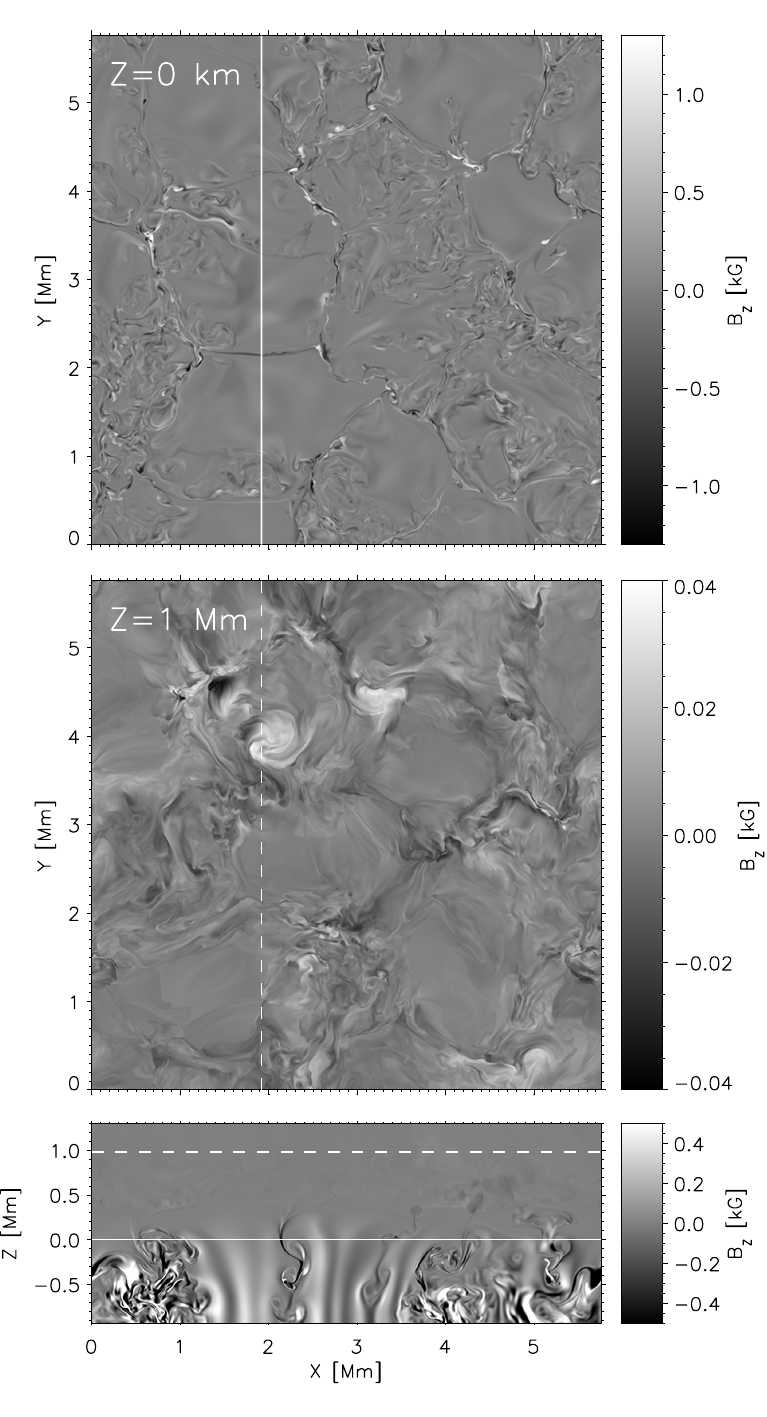}
\includegraphics[keepaspectratio,width=6cm]{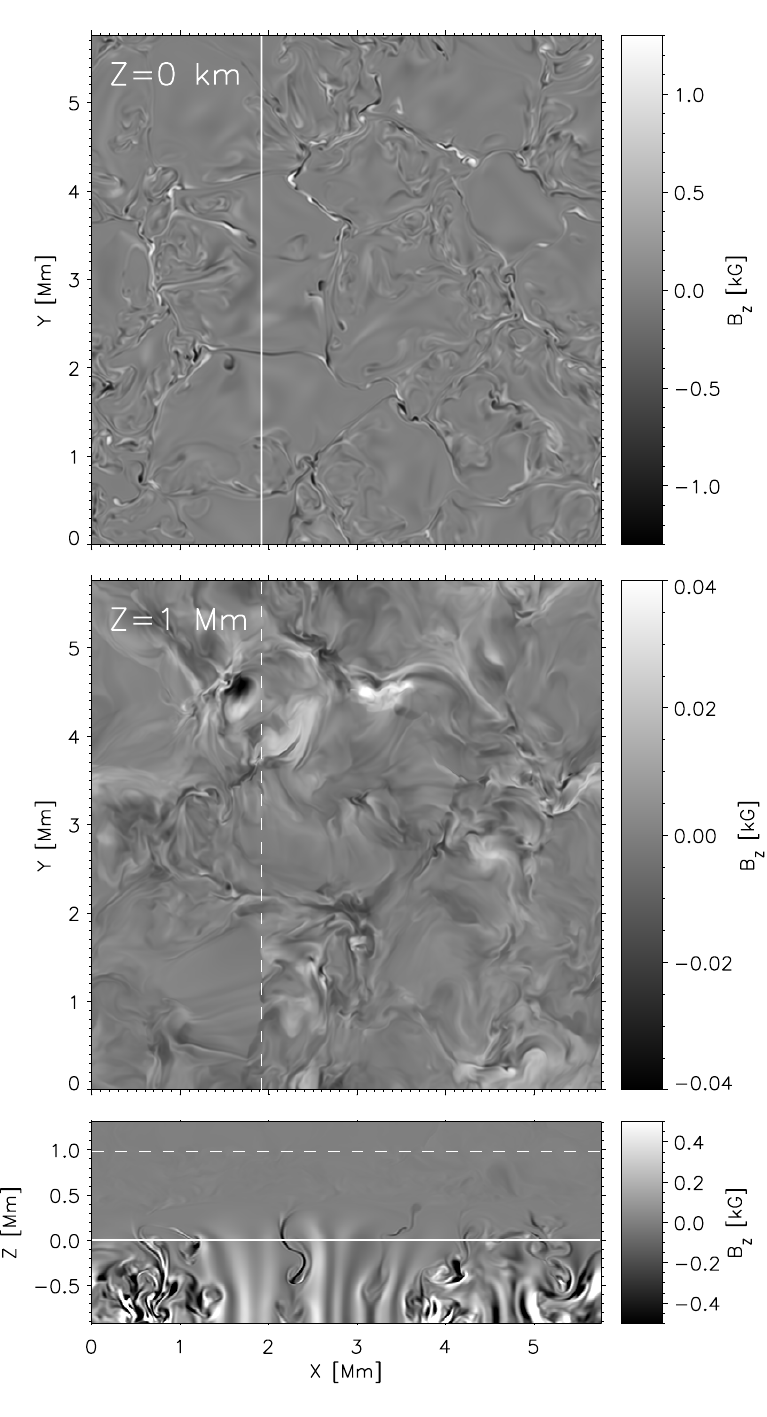}
\includegraphics[keepaspectratio,width=6cm]{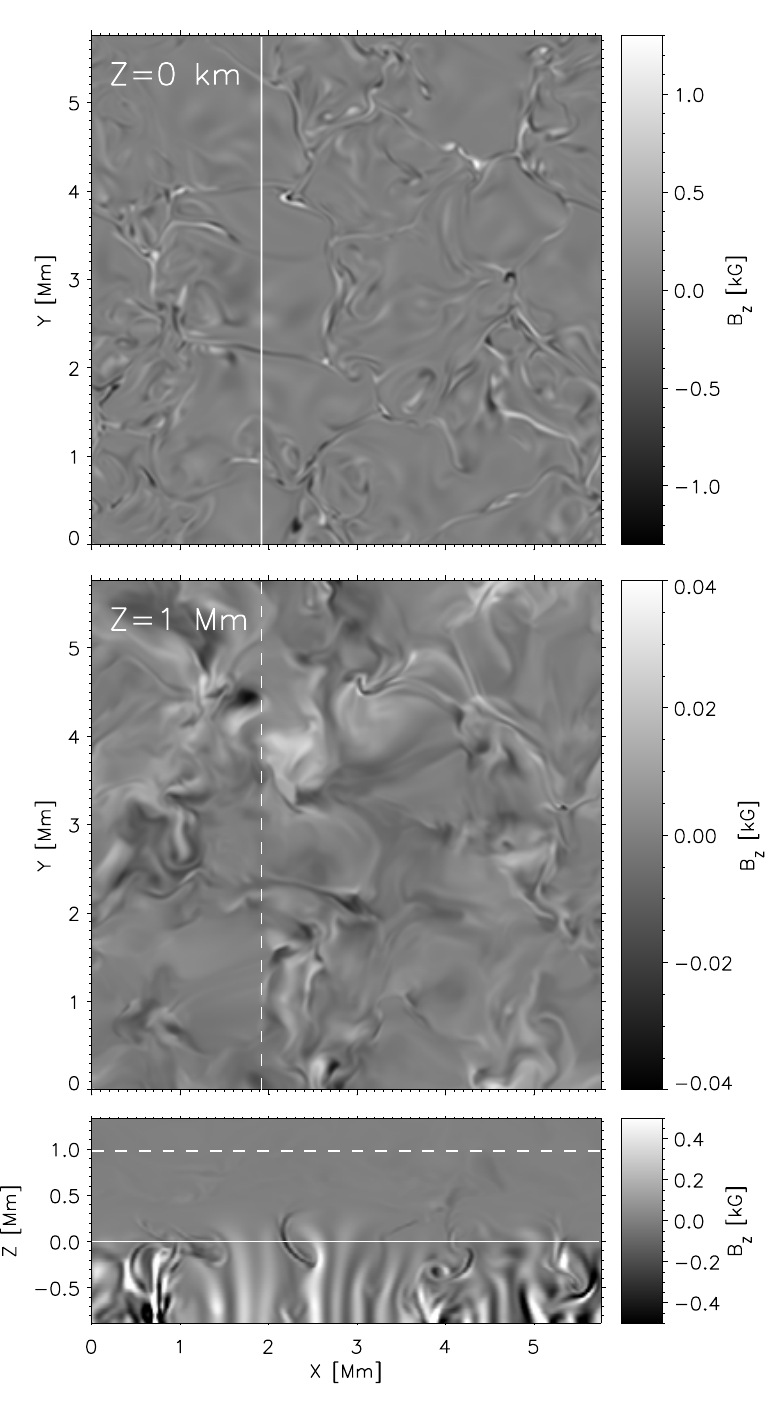}
\caption{\footnotesize Same as Fig.~\ref{fig:vzsnap-dyn} but for the vertical magnetic field component.}
\label{fig:bzsnap-dyn}
\end{figure*}
  
\begin{figure*}
\centering
\includegraphics[keepaspectratio,width=6cm]{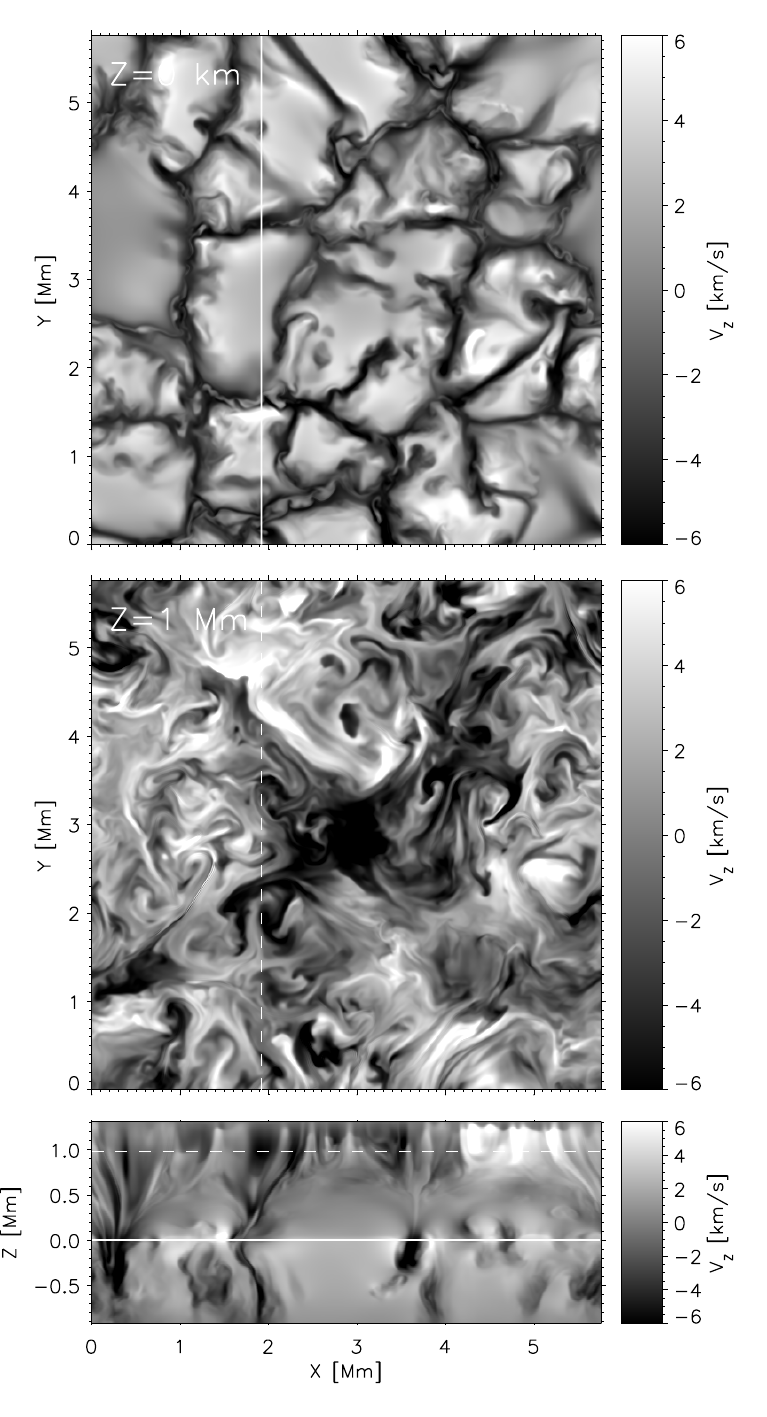}
\includegraphics[keepaspectratio,width=6cm]{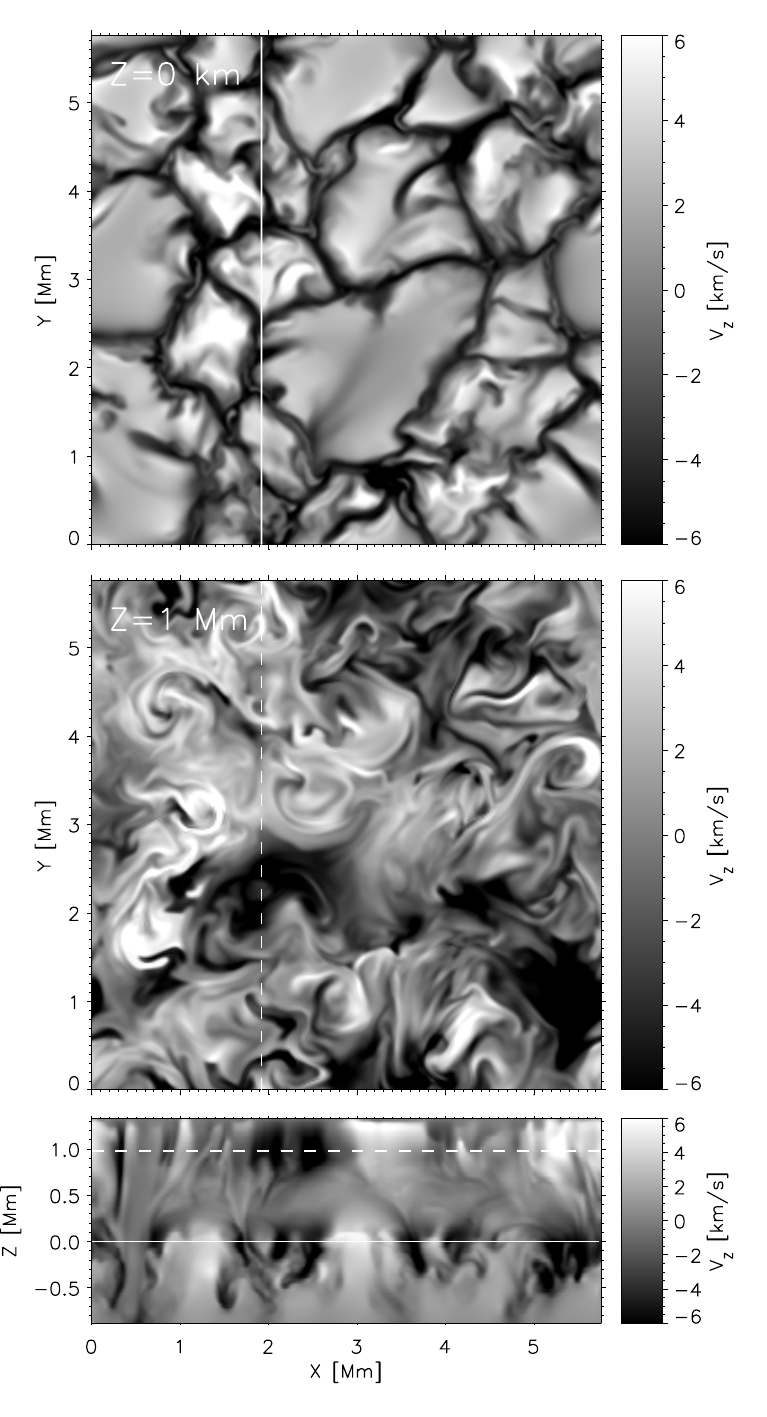}
\includegraphics[keepaspectratio,width=6cm]{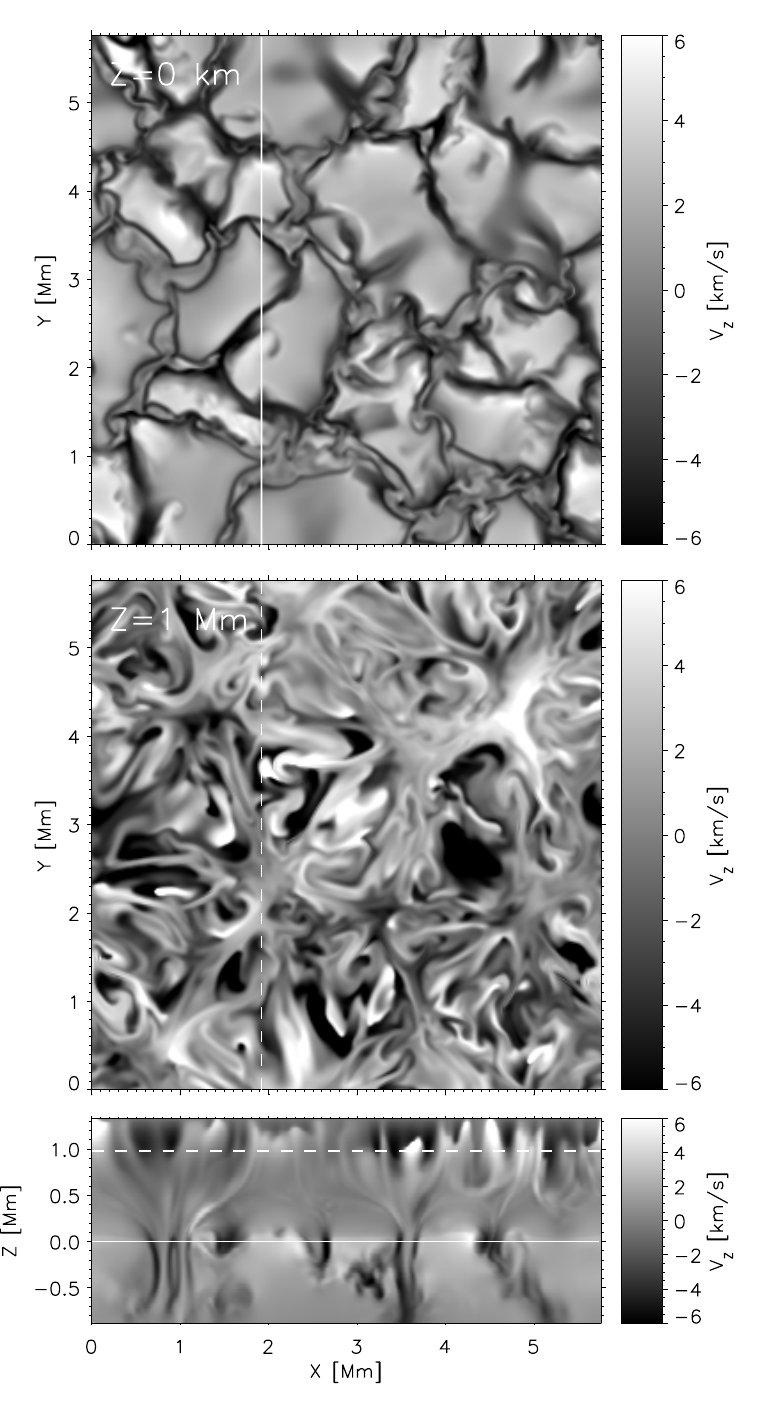}
\caption{\footnotesize Snapshots of the vertical velocities in the photosphere at $z=0$ km (top row), in the chromosphere at $z=1$ Mm (middle row) and in a vertical cut at an arbitrary horizontal location (bottom row). Horizontal lines at the top and middle panels mark the location of the slice from the bottom panel. Horizontal lines at the bottom panels mark the heights where the velocities at the upper panels are shown. Panels from left to right are for Bz50$_{10}$, Bz50$_{20}$ and Bz200$_{20}$ simulations, correspondingly.}
\label{fig:vzsnap-uni}
\vspace{6truemm}
\centering
\includegraphics[keepaspectratio,width=6cm]{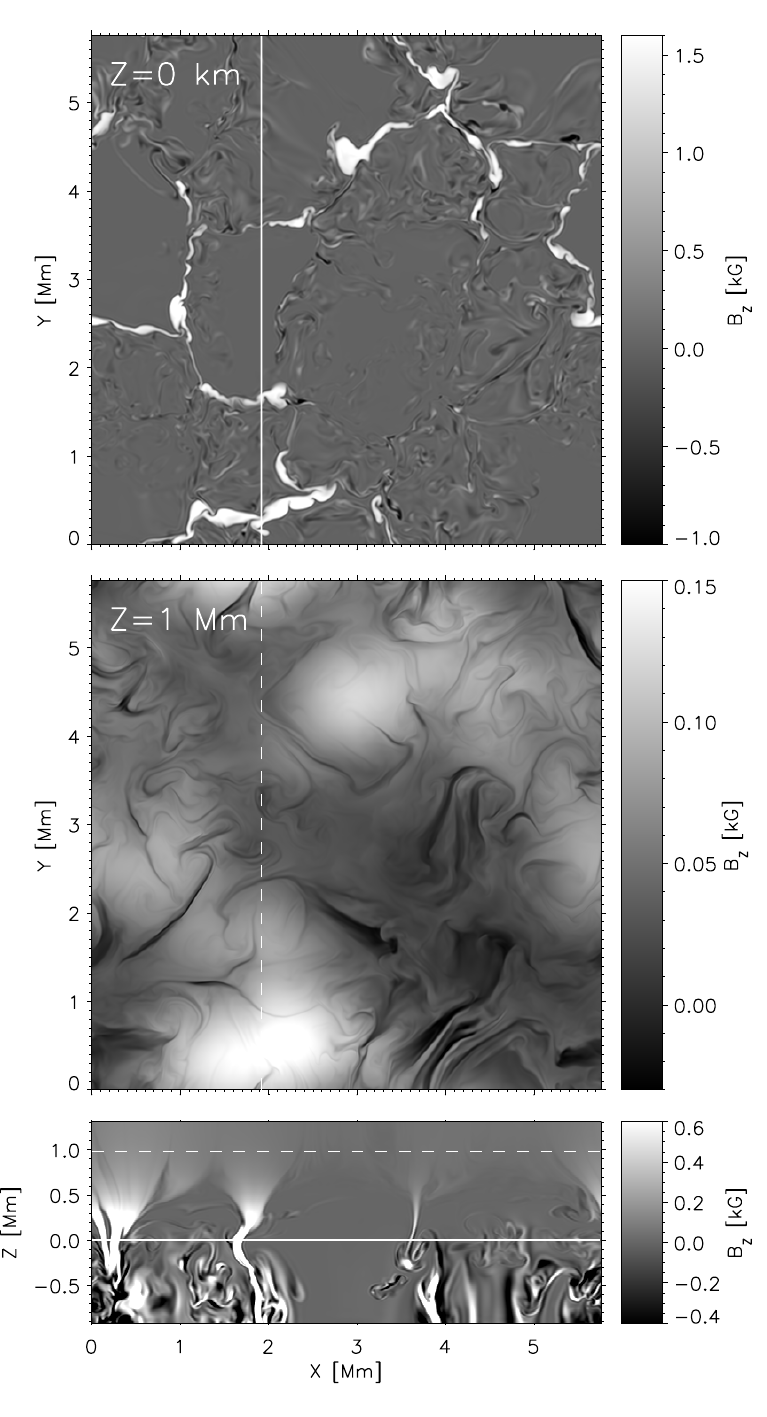}
\includegraphics[keepaspectratio,width=6cm]{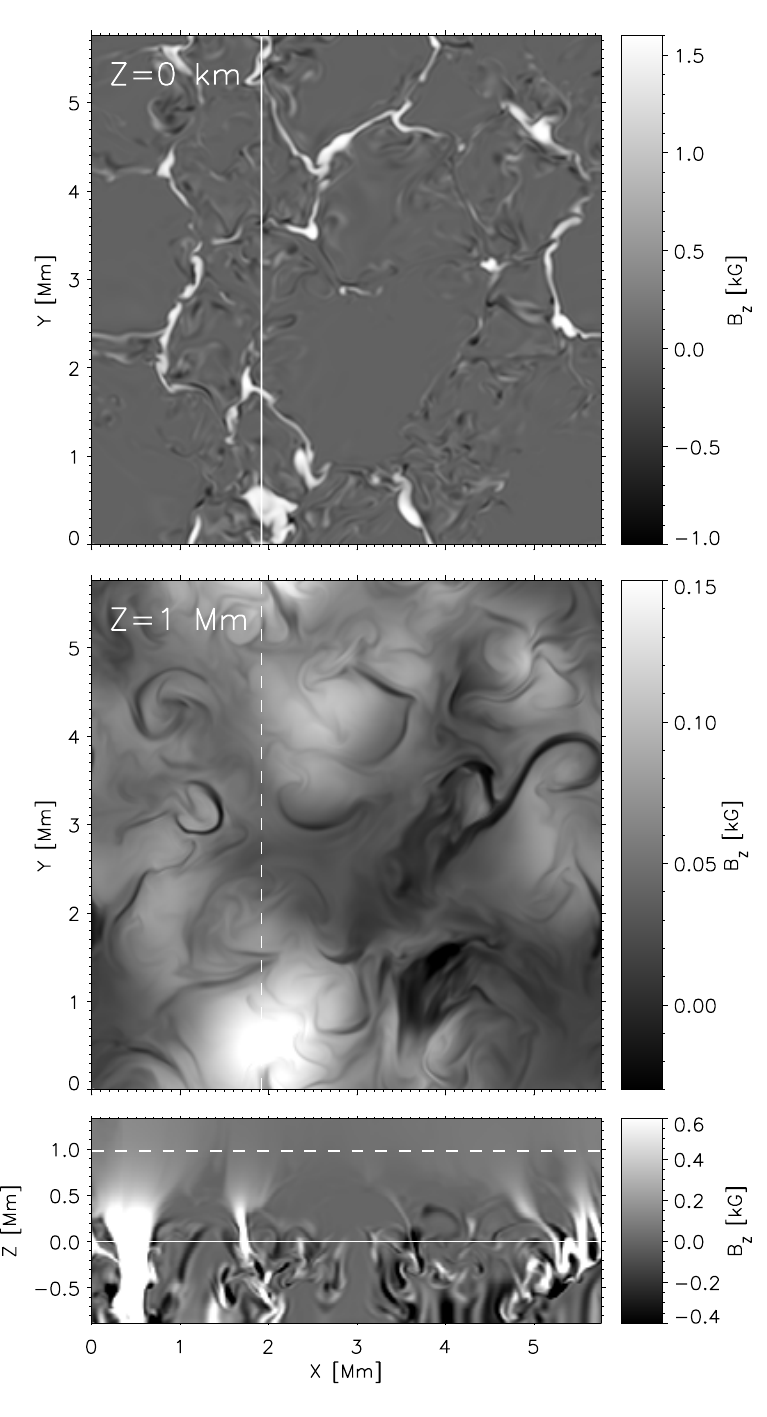}
\includegraphics[keepaspectratio,width=6cm]{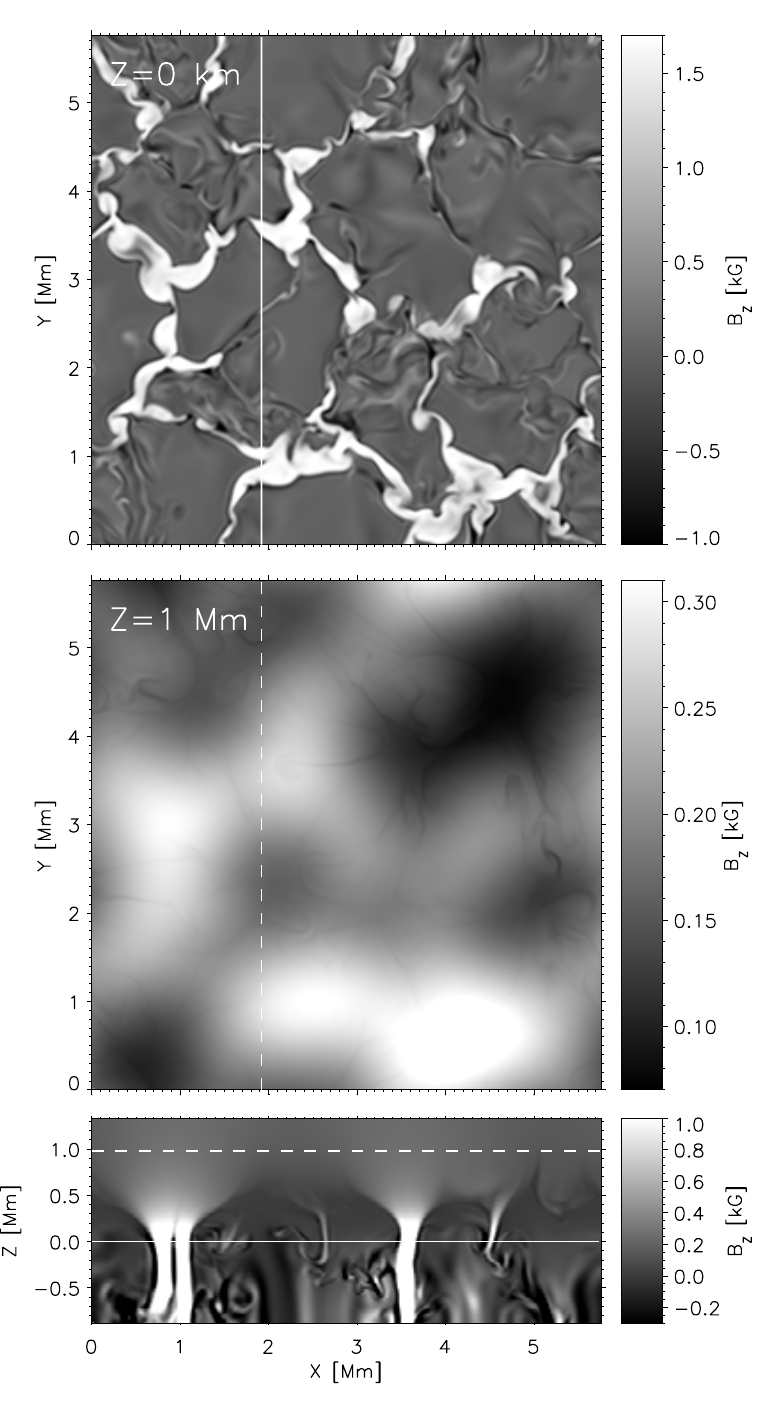}
\caption{\footnotesize Same as Fig.~\ref{fig:vzsnap-uni} but for the vertical magnetic field component.}
\label{fig:bzsnap-uni}
\end{figure*}

\section{Simulation setup}
\label{sect:setup}

Simulations were conducted using the \mancha code \citep{Modestov+etal2023}, and their configuration is comprehensively detailed in our prior research studies \citep{Khomenko+etal2018, Khomenko+etal2021, Gonzalez-Morales+etal2020}. Here we briefly summarize the setup of the simulations. \mancha solves the non-ideal non-linear equations of single-fluid magnetohydrodynamics (MHD), including the ambipolar diffusion, the Hall effect, and Biermann battery effect. The system of equations is closed with a realistic equation of state (EOS) for the solar chemical mixture given by \citet{1989GeCoA..53..197A} based on Saha equation. We use pre-computed tables to recover the gas and electron pressure and temperature given the internal energy and density as primary variables. The EOS takes into account the effects of the first and second atomic ionization for elements with atomic number below 92, and formation of hydrogen molecules. The local thermodynamic equilibrium (LTE) radiative losses are included here in gray approximation (wavelength-independent opacities). It needs to be kept in mind that the LTE approximation limits the accuracy of the radiative loss term in simulations extending to the chromosphere, therefore the vertical extent of our simulation box is limited to 1.4 Mm above the photosphere (see below). The horizontal boundaries are periodic. We used an open bottom boundary condition with mass and entropy controls that ensures that the models have the correct values of solar radiative flux. The bottom boundary has zero magnetic field inflow. This is done by setting the magnetic field to be vertical at the boundary (symmetric $B_z$ and antisymmetric $B_x$ and $B_y$ values in the ghost cells). The top boundary is closed for mass flows. We set symmetric boundary conditions (zero gradient) in internal energy and density variables at the top boundary and the temperature is computed through the EOS.

The simulation volume spans approximately 0.9 Mm below the photosphere to 1.4 Mm above it, encapsulating a physical box size of $5.76\times5.76\times2.3$ Mm$^3$ within the solar domain. The sole distinction between the simulations reported in \citet{Khomenko+etal2018, Khomenko+etal2021, Gonzalez-Morales+etal2020} and here lies in the alteration of spatial resolution.  Three distinct resolutions were employed for the simulations: $20\times20\times14$, $10\times10\times7$, and $5\times5\times3.5$ km$^3$, within boxes of dimensions $288\times288\times168$, $576\times576\times328$, and $1152\times1152\times648$ px$^3$, respectively. A schematic representation of the setup of our runs is depicted in Figure~\ref{fig:setup}. All simulations were initialized from a purely hydrodynamic initial condition with stationary developed convection at 20 km horizontal resolution. From there, we ran three different simulation series:  (Dyn) small-scale dynamo simulations with the magnetic fields initially seeded by the battery term; (Bz50) initially implanted vertical field of 50 G, typical of a medium-quiet solar region, and (Bz200) initially implanted vertical field typical for a solar network region. The latter two are indicated as `unipolar'' in the figure. For the Dyn runs, we ran the simulation for about 2 hours to reach the saturation regime of the small scale dynamo \citep[see][]{Khomenko+etal2017}.  In the case of the unipolar runs, the magnetic filed was evolved by the granular flow for approximately 30 solar minutes until a new stationary regime was reached. At this moment, the non-ideal effects (ambipolar, Hall) were introduced in each of the type of the simulations (Dyn, Bz50, Bz200) at the original 20 km resolution. Then, three series (``clean'', ambipolar, and ambipolar \& Hall) were run for a maximum of 20 minutes of solar time (with the exception of $Bz50_{20}$ case that was run for approximately 25 min). The ``clean'' label means that the simulation does not have any of the non-ideal effects. In the following, we will label these series as Dyn$_{20}$, Bz50$_{20}$, Bz200$_{20}$. In parallel, the resolution of the "clean" MHD snapshots was interpolated to either 5 or 10 km horizontally (and the corresponding sampling in vertical). These refined simulations were run for approximately 6 minutes of solar time to settle the new resolution. This stabilization of the convection led to different granular patterns in each resolution step that prevents from doing one to one comparison between the various series. 
Following this resolution refinement, the process of introducing non-ideal effects was repeated, and the simulations with enhanced resolution were conducted for up to 20 minutes of solar time, as illustrated in Figure \ref{fig:setup}. These series will be labeled as Dyn$_{10/5}$, Bz50$_{10/5}$, Bz200$_{10/5}$. The snapshots were saved every 10 sec.

According to our previous studies in idealized setups, already 10--15 solar minutes must be enough for the non-ideal effects to act. The characteristic time scales of the ambipolar diffusion in the chromosphere are comparable to the radiation cooling time scales, and are of the order of 200--300 sec, see \citet{Khomenko+Collados2012, Gonzalez-Morales+etal2020}. Since the structures do not significantly change during this time, these series will be used to perform a one to one comparison between the snapshots and to trace individual heating events. In the present study we only concentrate on the action of ambipolar diffusion (comparing the snapshots with and without this term at the same spatial resolution and originally implanted field conditions) and we do not further discuss the Hall effect. 

Table \ref{tab:setup} summarizes the duration of the simulation runs. Note that the most costly simulations with the highest magnetization and resolution have not been run for a long enough time (or at all) because of the insufficient computing time available to us.

\begin{figure*}
\centering
\includegraphics[keepaspectratio,width=16cm]{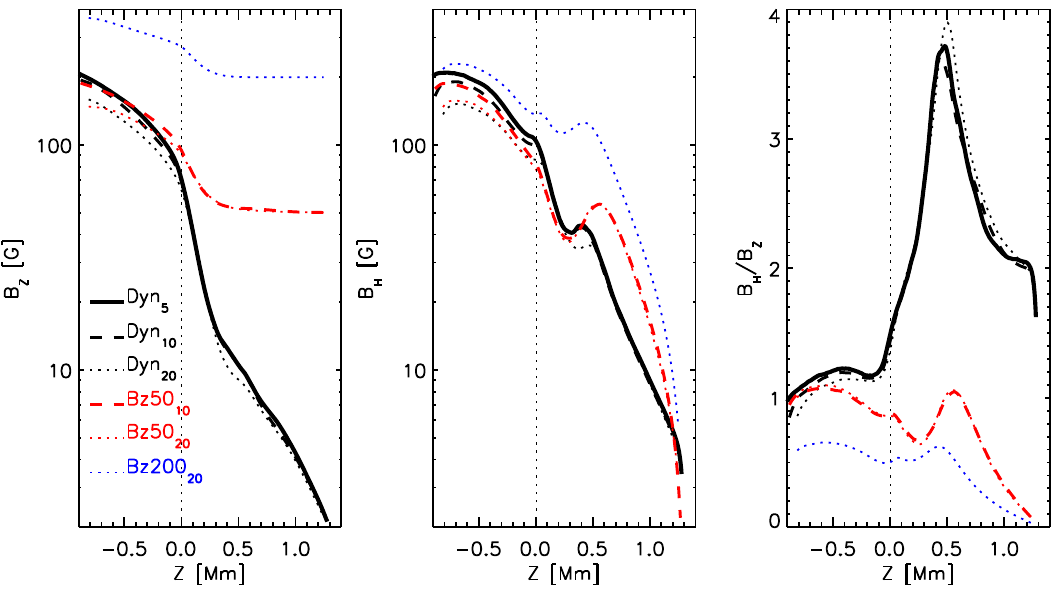}
\caption{\footnotesize Horizontally and temporally averaged magnetic field components in the different runs as a function of height. Left: vertical component, $\langle |B_Z|\rangle$, middle: horizontal component, $B_H=\langle \sqrt{B_X^2 +B_Y^2}\rangle$, right: ratio between the horizontal and the vertical components. Black lines are for Dyn model, red for Bz50 model and blue for Bz200 model. Solid thick line is for 5 km resolution, dashed medium-thick line is for 10 km resolution, dotted thin line is for 20 km resolution. }
\label{fig:bmean}
\end{figure*}

\section{Average properties of the simulations}
\label{sect:averages}

Figures \ref{fig:vzsnap-dyn}--\ref{fig:bzsnap-uni} depict snapshots illustrating the vertical velocity (positive values represent upward motions) and magnetic field at various heights within the solar atmosphere: bottom photosphere at $z=0$ km (top), low chromosphere at $z=1$ Mm (middle), and through all height range at a random horizontal location (bottom panel). Figures \ref{fig:vzsnap-dyn}--\ref{fig:bzsnap-dyn} do so for the dynamo simulations at three different spatial resolutions and the same moment in time. Figures \ref{fig:vzsnap-uni}--\ref{fig:bzsnap-uni} illustrate the same for the unipolar simulations, where for the Bz50 case the time moments are also the same for the two resolutions. These visual representations serve to show how the flow patterns and magnetic configurations undergo transformations in response to changes in resolution and magnetization. The photospheric region reveals that the intergranular regions host the bulk of turbulence, while the granules themselves exhibit a more laminar flows. This turbulence is visually better resolved in the higher resolution simulations, compare Dyn$_5$ case (Fig. \ref{fig:vzsnap-dyn} upper left) to the Dyn$_{20}$ case (Fig. \ref{fig:vzsnap-dyn} upper right). 

As we move into the chromosphere, the dynamics of the flow becomes significantly influenced by the magnetic topology. In the Dyn cases, the velocity field at $z=1$ Mm consists of an interference pattern of acoustic shocks, traveling horizontally through the domain. Meanwhile, the magnetic field exerts a notably weak influence, and the majority of the region is characterized by a large plasma beta. The turbulence in these shocks at 5 km resolution is visually much better resolved compared to our previous simulations \citep{Khomenko+etal2018}, and to other resolutions. The magnetic field in the Dyn$_5$ case shows a typical salt and pepper pattern characteristic for the small scale dynamo. In the chromosphere, these magnetic structures begin to expand, and larger scale concentrations can be observed. The time evolution reveals a lot of vorticity arising from the interaction of shocks, and the faint magnetic structures that rotate with the flow.  One of the instances of a vortical structure in magnetic field can be observed at the upper left part of the domain at the chromospheric slices, middle panels of Fig. \ref{fig:bzsnap-dyn}, around $x=2$ Mm and $y=4$ Mm. Note that this visual appearance of the rotation is clearly visible at 5 km resolution, but is much less evident at the coarser resolutions. The magnetic field has mixed polarity and is turbulent at the granular-intergranular borders. This turbulence extends over the granular zones in the higher resolution runs, becoming especially evident in the 5 km case (upper left panel of Fig. \ref{fig:bzsnap-dyn}).

In the Bz50$_{10}$ case, the flow in the chromosphere represents a flower-like structure in a strong correlation with the magnetic field (Figs.~\ref{fig:vzsnap-uni} and \ref{fig:bzsnap-uni}, left and middle panels). The velocity field at $z=1$ Mm is dominated by a mixture of field-aligned acoustic shocks and fast magnetic waves traveling across the field. A striking feature is the presence of distinct elongated dark zones within the magnetic field image, nearly devoid of a significant vertical magnetic component or exhibiting an opposite polarity. These areas align with regions of large upflow velocities and increased plasma densities. Unlike the dynamo case, the magnetic field in the unipolar cases has a dominant polarity, these unipolar strong magnetic concentrations are anchored into intergranular lanes. Their presence correlates with elongated features inside intergranular lanes at the photospheric level (upper panels of Figs.~\ref{fig:vzsnap-uni}, especially evident in the Bz200$_{20}$ case). The more resolved case shows additionally a turbulent mixed polarity component extended over a large fraction of granular regions, similarly to what was observed in the dynamo case.

A common feature across all cases is the decline in the amplitude of velocity fluctuations within the height range of roughly 0 to 800 km, followed by a subsequent increase beyond this height. This distinctive behavior, frequently encountered in various simulations, is a well-documented phenomenon \citep[for example, see][]{Fleck2021}. It arises from changes in the mechanism of energy transport as we ascend through the solar atmosphere. Convective energy transport stops at the surface, and the associated flows overshoot till approximately middle-photosphere heights \citep{Kostik2009}. Thus, the amplitudes of the granular flows decrease with height. However, convection excites waves and these waves grow in amplitude from the surface upwards. This amplification arises from the conservation of mechanical energy, although it is influenced by dissipative mechanisms. While these waves exhibit small amplitudes within the photosphere, they become more pronounced within the chromosphere, eventually becoming discernible in representations such as the one shown in Figures \ref{fig:vzsnap-dyn}, \ref{fig:vzsnap-uni}, bottom panels. It is interesting to highlight that the presence of strong magnetic field serves to create a more coherent linkage between the layers in velocity images. For instance, in the Bz200 and even the Bz50 simulations, flows can be traced from below the surface up into the chromosphere. However, this connectivity appears to be less pronounced in the Dyn$_5$ scenario, where flows between these height levels seem somewhat disjointed.

In the case of Bz200$_{20}$ (as depicted in Figs.~\ref{fig:vzsnap-uni} and \ref{fig:bzsnap-uni}, right panels), the photospheric flow and magnetic field patterns have similar structure to the Bz50$_{20}$, but with stronger magnetic field concentrations and even more magnetic field shaped flows. Unlike the lower magnetization cases, a distinct flow pattern linked to intergranular bright points in the photosphere becomes markedly evident. Upon entering the chromosphere, the velocity patterns adopt an even more pronounced flower-like structure. The level of small-scale details in the photospheric images is similar for 20 km resolution across all three magnetic field models. The chromospheric flows also show a similar level of details between  Bz200$_{20}$ and Bz50$_{20}$ cases. Nevertheless, the chromospheric magnetic field image exhibits notably diffused structures. In the 200 G case, the magnetic field exerts a more dominant influence, effectively removing the turbulence around magnetic field concentrations in the low plasma $\beta$ regions of the domain. The dark elongated lanes in the magnetic field image remain discernible. The magnetic field does not overturn, through, withing these dark lanes, and it generally maintains the same polarity through the chromosphere. The flux tubes expand with height and, since the field is stronger in the Bz200$_{20}$ case, these flux tubes are more visible and volume-filling in the upper chromospheric layers. Higher up the flux tubes merge and the field becomes relatively homogeneous, in sharp contrast with the velocity field.

\begin{figure*}
\centering
\includegraphics[keepaspectratio,width=16cm]{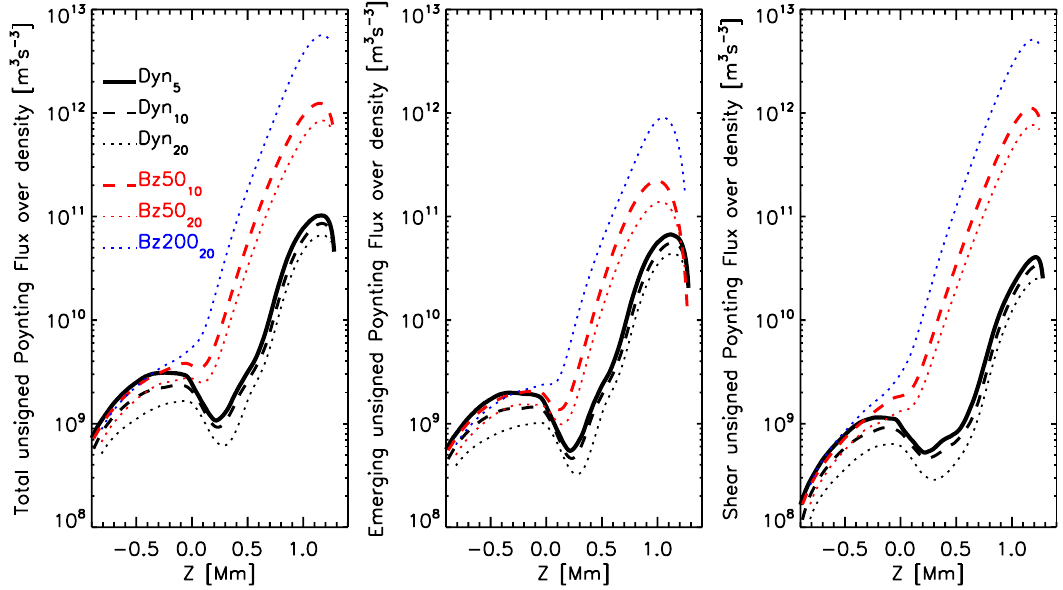}
\includegraphics[keepaspectratio,width=16cm]{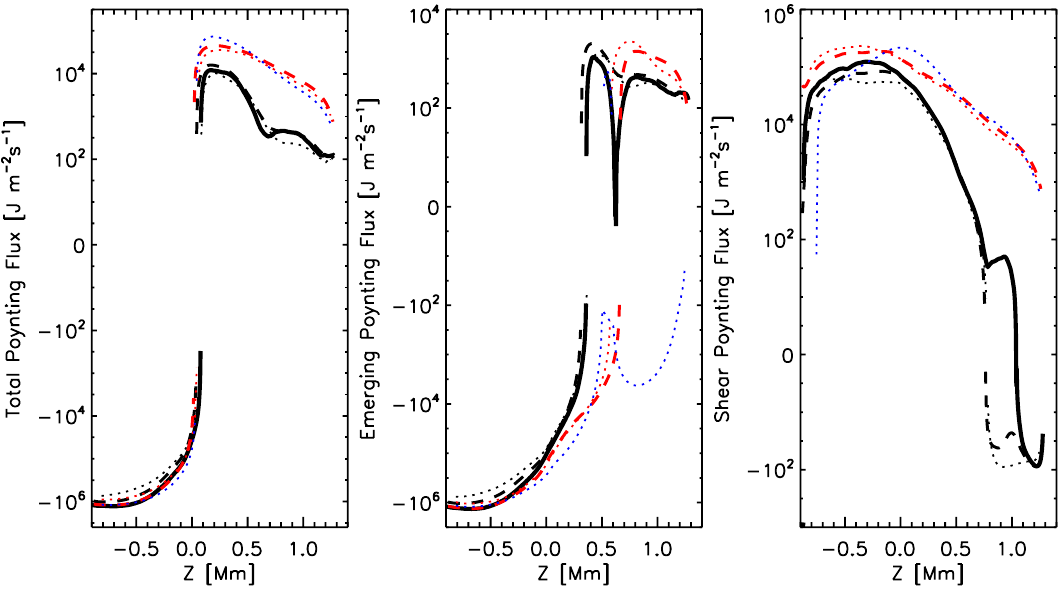}
\caption{\footnotesize  Upper panels, left: horizontally and time average of the unsigned vertical component of the magnetic Poynting flux in different runs as a function of height. For better visualization, the Poynting flux has been normalized over the density. Middle: emerging Poynting flux, vertical component, computed as $\langle S_{\rm EM}^{\rm emer}\rangle=\langle |v_z(B_x^2+B_y^2)/\mu_0| \rangle$. Right: shear Poynting flux, vertical component, computed as $\langle S_{\rm EM}^{\rm shear}\rangle=\langle |-B_z(v_xB_x + v_yB_y)/\mu_0|\rangle$.  Black lines are for Dyn model, red for Bz50 model and blue for Bz200 model. Solid thick line is for 5 km resolution, dashed medium-thick line is for 10 km resolution, dotted thin line is for 20 km resolution. Bottom panels represent the same, but without dividing by density and taking into account the sign while averaging. Negative values mean downward propagating flux. The apparent discontinuity in the curves is due to representation, caused by the change of sign.}
\label{fig:pfmean_overrho}
\end{figure*}

\subsection{Magnetic field strength}

Figure \ref{fig:bmean} shows how the average magnetic field scales with the resolution and the type of the model. In the dynamo simulations (black lines) the magnetic field at all layers increases from 20 to 10 and 5 km resolution. This increase is more pronounced in the subsurface layers. The average field strength at $z=0$ is $|B|=$128, 120 and 107 G for the Dyn$_5$, Dyn$_{10}$ and Dyn$_{20}$ km simulations, respectively.  It is interesting to note that the difference in the field strength between Dyn$_{20}$ and Dyn$_{10}$ km resolutions is larger than between Dyn$_{10}$ and Dyn$_{5}$ km resolutions, indicating that the dynamo is saturating. Both vertical (left panel) and horizontal (middle panel) field strengths in the dynamo run scale with the resolution. The scaling of the average magnetic field with the resolution in the dynamo simulation can be anticipated since the Reynolds number is expected to increase with resolution so that the dynamo action becomes more efficient. This is verified by our calculations, see Section~\ref{sect:heating} below.

In all Dyn simulations, the vertical magnetic field smoothly decreases with height till about 1 G in the chromosphere. On the contrary, the horizontal component has a local maximum around 0.5 Mm, related to the expansion of the magnetic field structures. Above $z=0$, through the photosphere and chromosphere, the horizontal field is always larger than the vertical one, being most horizontal at about \hbox{0.5 Mm} height.  There is a weak preference for the vertical field below $z=0$, keeping in mind that an isotropic field would mean $B_H=\sqrt{2}B_z$. The return to the vertical field higher up is due to our boundary condition that requires vertical field at the upper boundary. Nevertheless, the field there is very weak. By looking at the deviations with resolution from this general behavior, it is seen that the magnetic field is more horizontal on average in the deeper layers and more vertical in the upper layers with the highest resolution, though the reason for that is not clear.

The models with implanted field have a different behavior (red and blue curves in Fig.~\ref{fig:bmean}). The scaling with resolution in the 50 G case is only evident in the subsurface layers. Higher up neither vertical nor horizontal fields change with the resolution. The vertical field converges to a constant value of 50 G in the upper layers, following the magnetic flux conservation, while the horizontal component keeps decreasing. The horizontal component has a local maximum around 0.5-0-7 Mm height, depending on the average magnetization. The field expansion happens deeper in the 200 G case compared to the 50 G case. We do not observe scaling with the resolution for the ratio of the horizontal to vertical field in the 50 G case.

\subsection{Poynting flux}

The overall conclusion from Fig.~\ref{fig:bmean} is that, on average, the magnetic energy contained in the simulations scales with the resolution. This has repercussions into the propagation of the magnetic Poynting flux (given by Eq. \ref{Poyntingvector}). 

\begin{equation}
\mathbf{S}_{\rm EM}=-\frac{(\mathbf{v} \times \mathbf{B}) \times \mathbf{B}}{\mu_0},
\label{Poyntingvector}
\end{equation}
Here $\mathbf{v}$ is the velocity, $\mathbf{B}$ is the magnetic field and $\mu_0$ is magnetic permeability constant.

In Figure \ref{fig:pfmean_overrho}, we show the vertical component of the total ideal Poynting flux (left panels), together with its  ``emerging'' (central panels) and ``shear'' parts (right panels), following  \cite{Shelyag2012},

\begin{eqnarray} \label{eq:poynting_ideal}
S_{\rm EM}^{\rm emer}&=&v_z\cdot(B_x^2 + B_y^2)/\mu_0; \\ 
S_{\rm EM}^{\rm shear}&=&-B_z\cdot(B_x v_x + B_y v_y)/\mu_0. 
\end{eqnarray}
The ``shear'' part of the flux corresponds to horizontal motions along vertical flux tubes, while the ``emerging'' part corresponds to horizontal magnetic field perturbations transported by vertical plasma motions. The relative amount of magnetic energy between the simulations is better displayed by plotting the quantity of $\mathbf{S}_{\rm EM}/\rho$, where $\rho$ is the density. The upper panels of the figure shows the space and time averaged absolute values of $\langle|S_{\rm EM}/\rho|\rangle$, $\langle|S_{\rm EM}^{\rm emer}/\rho|\rangle$ and $\langle|S_{\rm EM}^{\rm shear}/\rho|\rangle$. According to the figure, the fluxes, normalized by density, exhibit a non-monotonic dependence on height. The flux reaches a minimum around the middle-photosphere, reflecting once again the transition of the transport mechanism from convection to waves. This transition occurs at slightly different heights in the dynamo and unipolar simulations. Another aspect illustrated by the figure is that all the considered quantities scale with resolution: the amount of the available Poynting flux increases with increasing resolution. In the atmospheric layers of the model, the flux is largest in simulations with higher magnetization, specifically Bz200$_{20}$. Another aspect to note is that the unipolar models are dominated by the shear component of the flux, which naturally happens because the magnetic field tends to the vertical in the upper regions of the models. The situation is different in the dynamo models, where the transport by vertical plasma motions is slightly superior. Similar behavior was reported previously in \cite{Shelyag2012}.

Therefore, we conclude that, despite the simulations being made identical and only changing the numerical resolution, the amount of available magnetic energy scales with the resolution (by about 30-50 percent, depending on the model), reflecting the increase in the average magnetic field and the amplitude of plasma motions. Since ambipolar diffusion is supposed to act towards converting magnetic energy into thermal energy of the plasma, the above results suggest that a larger amount of magnetic energy is available for dissipation, and greater heating over the volume might be achieved in the models with higher resolution.

Bottom panels of Figure \ref{fig:pfmean_overrho} show a more standard representation of the Poynting fluxes, averaged taking into account the sign and without the division by density. The average total flux (left panel) is negative in the sub-surface layers, caused by the behavior of its emerging part, $S_{\rm EM}^{\rm emer}$. The latter is because the strong magnetic concentrations are preferentially located in down-flowing intergranular regions. The total flux is on average positive in the atmospheric part of the domain. There, the scaling with resolution is not so evident, though one can trace a slight dominance of the flux in the models with higher resolution for each magnetization. More flux is transported on average in the unipolar models, caused by the dominance of the `shear'' flux $S_{\rm EM}^{\rm shear}$. Interestingly, Bz200$_{20}$ case shows values of the total $S_{\rm EM}$ slightly below those in Bz50$_{20}$. This is probably caused by the suppression of fluctuations thanks to the strong magnetic field in the 200 G model, which was already seen in its more diffused appearance in Fig.~\ref{fig:bzsnap-uni}. The emerging component (middle panel) presents a non-monotonic behavior in the atmospheric regions with a depression, and even negative values around 0.5 Mm.  These negative values are most probably a drawback of the insufficient averaging since the Dyn$_5$ case was run for only a short amount of time. This can be traced back to Fig.~\ref{fig:bmean} at heights where the field opens up and becomes more horizontal. Such a behavior changes the nature of propagating waves and probably causes additional reflections.

\section{Heating mechanisms}
\label{sect:heating}

There are several sources of dissipation and heating in our simulations. On the one hand, the series with ambipolar diffusion contain a physical heating mechanism based on this effect. This results in an additional term in the internal energy equation, proportional to the ambipolar diffusion coefficient, $\eta_{\rm A}$, and to the currents perpendicular to the magnetic field, $J_{\perp}$,
\begin{equation} \label{eq:ambidiff}
\frac{\partial e_{\rm int}}{\partial t} + (...) = 
\mu_0\eta_{\mathrm A}{J_\perp}^2,
\end{equation}
where $e_{\rm int}$ stands for internal energy. The $\eta_{\rm A}$ coefficient is defined as,
\begin{equation} \label{eq:etaa}
\eta_{\mathrm A} =\frac{\xi_{\mathrm n}^2\mathbf{B}^2}{\alpha_{\mathrm n}\mu_0}.
\end{equation}
In this expression $\xi_{\rm n}$ is mass fraction of neutrals, and $\alpha_{\rm n}$ is the neutral collisional parameter, which depends on the partial number densities and collisional frequencies between the plasma components \citep{Khomenko+etal2014}. The units of $\eta_{\mathrm A}$ are [m$^2$s$^{-1}$]. The current perpendicular to the magnetic field is defined as,
\begin{equation}
\label{eq:jota}
\mathbf{J}_\perp = -\frac{(\mathbf{J}\times\mathbf{B})\times\mathbf{B}}{\mathbf B^2}.
\end{equation}

\noindent On the other hand, there are terms related to the numerical dissipation. \mancha code works with two numerical stabilization techniques that prevent energy from building up at the smallest scales, unresolved by the grid. These are hyper-diffusion and filtering \citep{Modestov+etal2023}. The hyper-diffusion terms are intended to mimic to some extent the physical viscosity and Ohmic diffusivity. The application of the filtering is equivalent to adding an additional dissipation term to the equations, proportional to $\nabla^6 u$, i.e. to the 6$^\mathrm{th}$ derivative of a variable $u$. 
\begin{equation}
\label{eq:filt}
\left(\frac{\partial u}{\partial t}\right)_{\rm filt}=\nu^F_6\nabla^6 u.
\end{equation}   
The proportionality coefficient, $\nu^F_6$, depends on the time interval at which the filtering is applied, $\Delta t_{\rm filt}$, as it is computed as,
\begin{equation}
\label{eq:filtcoef}
\nu^F_6=\frac{\Delta s^6}{64\Delta t_{\rm filt}},
\end{equation}
where $\Delta s$ is the grid size. The filtering is done independently in three Cartesian directions, so that $\Delta s=\{\Delta x, \Delta y, \Delta z\}$, and the value of $\nu^F_6$ in general depends on the direction. The units of $\nu^F_6$ are [m$^6$s$^{-1}$]. \cite{Popescu+etal2023} performed a numerical check of the validity of Eqs.~\ref{eq:filt} and \ref{eq:filtcoef} in a case where filtering was the only stabilizing mechanism of their simulations, and confirmed that Eq.~\ref{eq:filtcoef} gives precise enough values of the numerical dissipation due to filtering. Note that an operator like the one in Eq.~\ref{eq:filt} is frequently referred to as hyper-diffusive operator \citep[see for example][]{Brandenburg2003, Maron+MacLow2009}. Nevertheless here we prefer not to use this nomenclature, both for consistency with our previous publications and also because technically in the code we do not apply Eq.~\ref{eq:filt} but perform an equivalent sixth-order digital filter following \cite{Parchevsky+Kosovichev2007}.

\begin{figure}
\centering
\includegraphics[keepaspectratio,width=7cm]{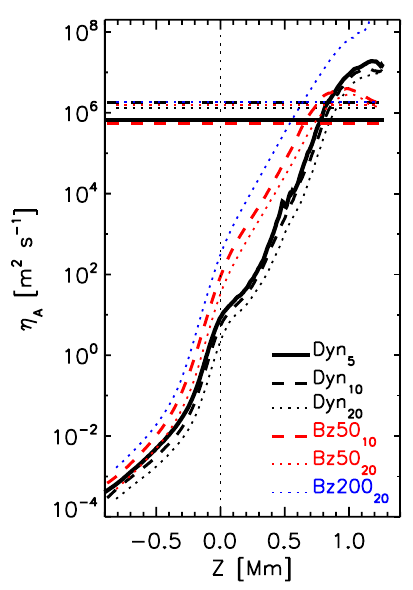}
\caption{\footnotesize Horizontally and time average values of the ambipolar coefficient, $\eta_A$, together with the artificial diffusion, $\nu_6^F/ds^4$ (horizontal lines), obtained after the fit of Eq.~\ref{eq:nufit}. The meaning of the curves is indicated in the figure. }
\label{fig:ambi}
\end{figure}

In the current simulations we set to zero the Ohmic hyper-diffusive terms both in the induction and in the energy balance equations. We kept, though, the hyper-diffusive terms in the rest of the equations, see \citet{Modestov+etal2023}, section 3.3. The reason for this choice is to minimize the fictitious numerical diffusion and to maximize the effect of the physical ambipolar diffusion. This choice is also consistent with our previous studies \citep{Khomenko+etal2017, Gonzalez-Morales+etal2020} and it leads to the small-scale dynamo experiments to saturate at values of the magnetic field strength very similar to observations \citep{TrujilloBueno2004}.

Altogether with the filtering terms, this brings to the following form of the diffusive terms in the momentum, induction and the internal energy equations,
\begin{equation} \label{eq:vdiff}
\left( \frac{\partial \rho \mathbf{v}}{\partial t}\right)_\mathrm{diff} =\nabla \cdot \boldsymbol{\tau} + \nu^F_6\nabla^6 \rho\mathbf{v}, 
\end{equation}
\begin{equation} \label{eq:bdiff}
\left( \frac{\partial  \mathbf{B}}{\partial t}\right)_\mathrm{diff} = \nu^F_6\nabla^6 \mathbf{B},
\end{equation}
\begin{equation} \label{eq:ediff}
\left(\frac{\partial e_{\rm int}}{\partial t}\right)_{\mathrm{diff}} = \boldsymbol{\tau} : \nabla \mathbf{v} + Q^{\rm MAG}_{\rm filt} + Q^{\rm VISC}_{\rm filt},
\end{equation}
where : stands for tensor double contraction;  $\boldsymbol{\tau}$ is the viscous stress tensor with components: 
\begin{equation}\label{eq:tau}
\tau_{ij} =\frac{1}{2}\rho\,\left(\nu_{j} (v_{i})\,\frac{\partial v_{i}}{\partial x_j} + \nu_{i} (v_{j})\,\frac{\partial v_{j}}{\partial x_i}\right),
\end{equation}
with $\nu_{j} (v_{i})$ being numerical diffusion coefficients, computed following \citet{Modestov+etal2023}, see their Section 3.4. The units of $\nu_{j} (v_{i})$  are [m$^2$s$^{-1}$]. 

The equivalent viscous and magnetic heating terms by the 6th order operator can be derived from the momentum and induction equations:

\begin{equation} \label{eq:vv6}
\frac{1}{2}\left( \frac{\partial \rho \mathbf{v}^2}{\partial t}\right)_\mathrm{diff} = \nu^F_6{\mathbf v}\cdot\nabla^6 {\rho\mathbf v}.
\end{equation}

\begin{equation}\label{eq:bb6}
\frac{1}{2\mu_0}\left( \frac{\partial  \mathbf{B}^2}{\partial t}\right)_\mathrm{diff} =-\mu_0\nu^F_6 {\mathbf J}\cdot \nabla^4 {\mathbf J} = \frac{\nu^F_6}{\mu_0}{\mathbf B}\cdot\nabla^6 {\mathbf B},
\end{equation}
where the ``$\cdot$'' stands for the scalar product. The right-hand-side terms in the above equations have the same mathematical structure. The 6th order operator does not guarantee the positivity of Eqs.~\ref{eq:vv6}, \ref{eq:bb6}.  Note that the magnetic term is equivalent to a classical Joule heating term, but with an electric field proportional to $\nabla^4{\mathbf J}$ instead of simply ${\mathbf J}$. Since we are interested in the positive part, which would be equivalent to a heating, we further split the right-hand-side terms into an always positive component and a divergence of a flux. For Eq.~\ref{eq:bb6} it reads,

\begin{equation} \label{eq:slava}
    {\mathbf B}\cdot\nabla^6 {\mathbf B} = - (\nabla\nabla^2{\mathbf B} : \nabla\nabla^2{\mathbf B}) + \nabla\cdot \mathbf{F},
\end{equation}
where
\begin{equation}
    \mathbf{F} = (\nabla \nabla^4{\mathbf B})\cdot\mathbf{B} - (\nabla{\mathbf B})\cdot(\nabla^4{\mathbf B}) + (\nabla \nabla^2{\mathbf B})\cdot(\nabla^2{\mathbf B}),
\end{equation}
and similar for Eq.~\ref{eq:vv6}. Following \citet{Lukin+etal2024}, we approximate the heating through the positive term in Eq.~\ref{eq:slava} (first term on the right hand side), obtaining,

\begin{equation}
   Q^{\rm VISC}_{\rm filt} = \nu^F_6 \sum_{j=x,y,z}\left\{ \frac{1}{\rho}\left|\nabla[\nabla^2(\rho v_{j})]\right|^2 \right\}.
\end{equation}
and
\begin{equation}
\label{eq:qjoule_num}
Q^{\rm MAG}_{\rm filt}=
\frac{\nu^F_6}{\mu_0} \sum_{j=x,y,z}{\left|\nabla (\nabla^2 B_{j})\right|^2},
\end{equation}

\noindent In the discussion below we do not consider the contribution of the hyper-diffusive thermal conduction to the internal energy equation, though it is present in the calculation of the models. Since the thermal energy flux due to conduction can be both positive and negative, it does not represent a heating mechanism.

\begin{table}
  \caption{Theoretical (Th) and empirical (Em) diffusion coefficients in [m$^2$s$^{-1}$], $\nu_6^F/\Delta s^4$ (2nd and 3rd columns); approximate hydrodynamic and magnetic Reynolds numbers (4th and 5th columns). $\Delta s=\{\Delta x, \Delta y, \Delta z\}$ represents the grid size in the three Cartesian directions. The average value over the three directions is shown.} 
\begin{tabular}{ c | c | c | c | c}
  \hline \\
   Run & $\nu_6^F/ds^4$ (Th) & $\nu_6^F/ds^4$ (Em) & Re & Rm \\
   \hline \\
   Dyn$_5$      & 1.6$\times 10^6$ & 0.76$\times 10^6$ & 400 & 1300 \\ 
   Dyn$_{10}$   & 1.3$\times 10^6$ & 1.8$\times 10^6$  & 290 & 940  \\ 
   Dyn$_{20}$   & 2.6$\times 10^6$ & 1.3$\times 10^6$  & 160 & 730  \\ 
   Bz50$_{10}$  & 1.3$\times 10^6$ & 0.55$\times 10^6$ & 250 & 2700 \\ 
   Bz50$_{20}$  & 2.6$\times 10^6$ & 1.5$\times 10^6$  & 110 & 1300 \\ 
   Bz200$_{20}$ & 2.6$\times 10^6$ & 1.9$\times 10^6$  & 60  & 2800 \\ 
   \\ \hline 
  \end{tabular}
  \label{tab:diff}
\end{table}

\begin{figure*}
\centering
\includegraphics[keepaspectratio,width=16cm]{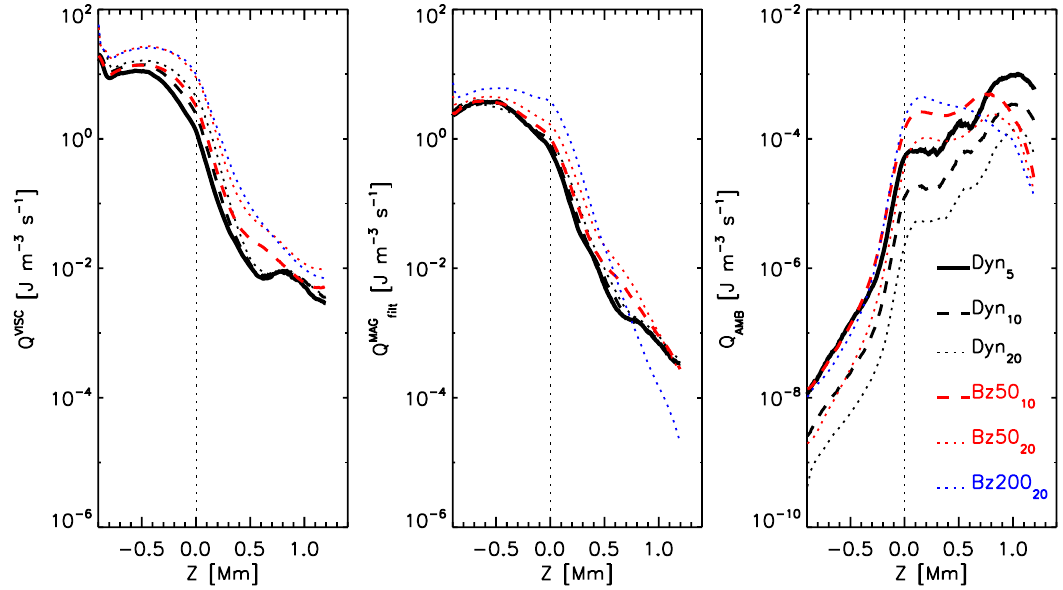}
\caption{\footnotesize Horizontally and time average values of the sum of viscous heating terms, $Q^{\rm VISC}_{\rm filt}+\boldsymbol{\tau} : \nabla \mathbf{v}$ (left); magnetic heating terms, $Q^{\rm MAG}_{\rm filt}$ (middle); and ambipolar heating term, Eq.~\ref{eq:ambidiff} (right), as functions of height. The meaning of the curves is indicated in the figure. Notice the difference in scale at the vertical axis. }
\label{fig:heating-z}
\end{figure*}

\begin{figure*}
\centering
\includegraphics[keepaspectratio,width=19cm]{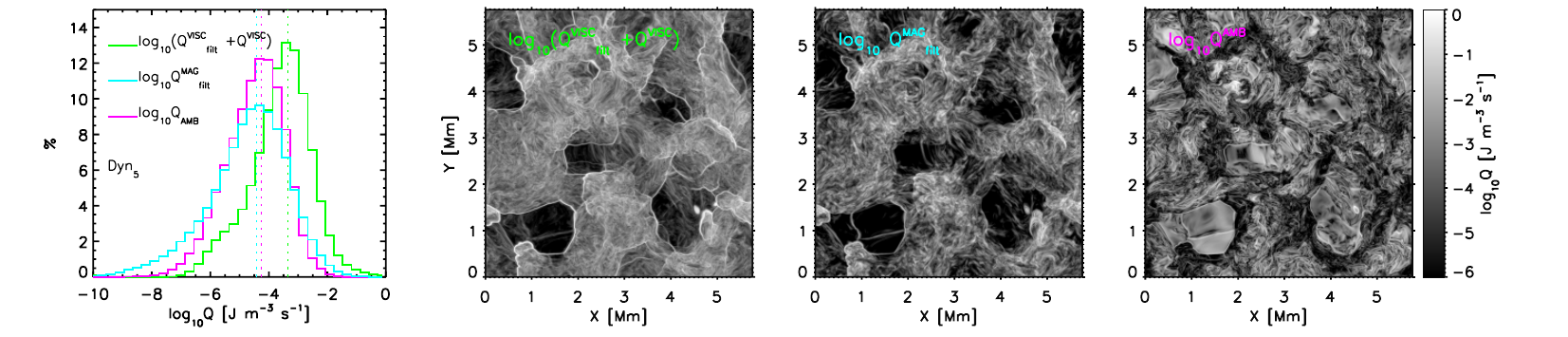}
\includegraphics[keepaspectratio,width=19cm]{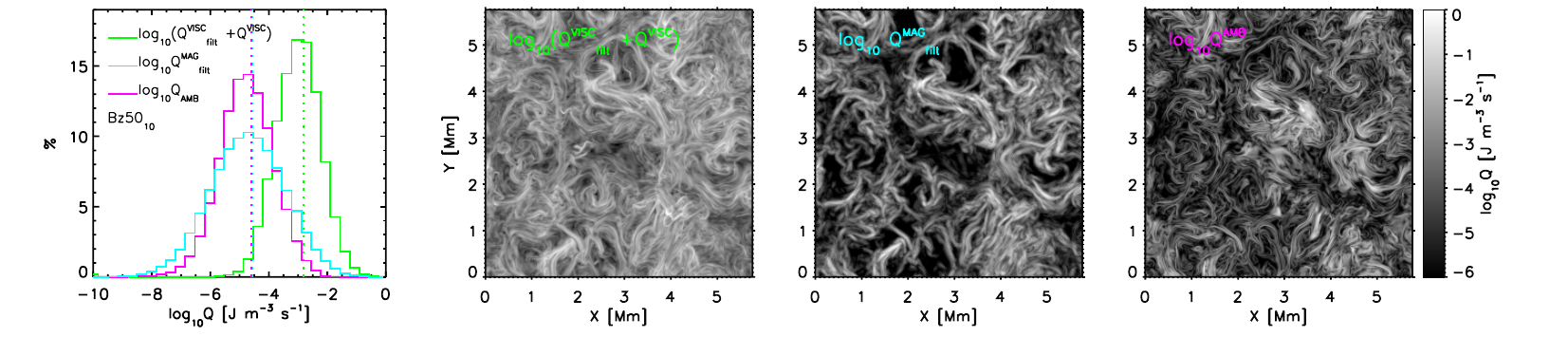}
\includegraphics[keepaspectratio,width=19cm]{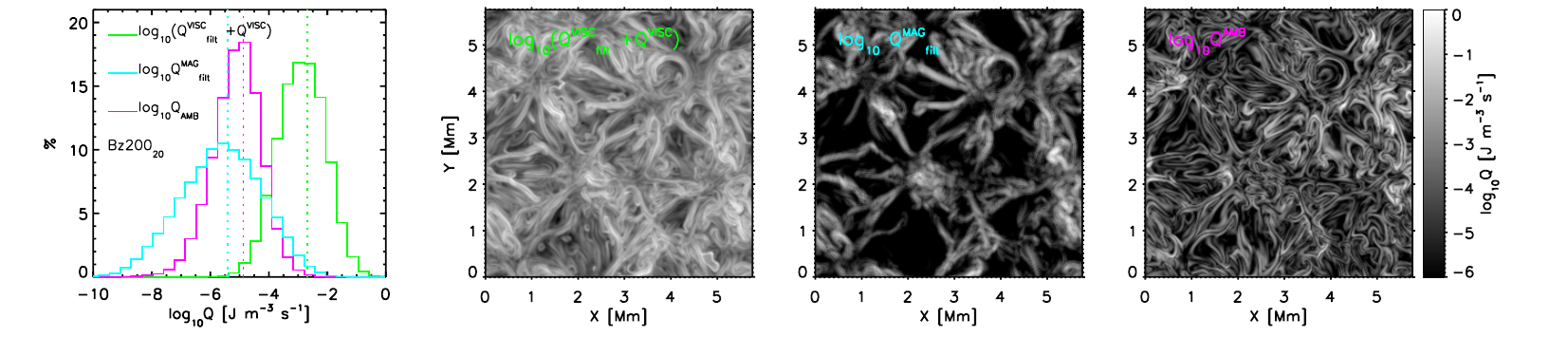}
\caption{\footnotesize Left column: histograms of the ambipolar (magenta), magnetic (light blue) and total viscous (light green) heating terms in the Dyn$_{5}$ (top), Bz50$_{10}$ (middle) and Bz200$_{20}$ (bottom) models. Vertical dotted lines mark the locations of the maxima of each distributions. Panels on the right show the horizontal distribution of the heating terms at 1 Mm height. The grayscale range is made the same in all the panels. }
\label{fig:heating-hist}
\end{figure*}

We are interested in verifying how the physical ambipolar diffusion, $\eta_A$, compares to the numerical magnetic diffusion due to filtering, $\nu_6^F$. For that, we computed $\nu_6^F$ both, following Eq.~\ref{eq:filtcoef} (Theoretical, Th, in Table  \ref{tab:diff}), and also repeating the procedure described in \cite{Khomenko+etal2017, Popescu+etal2023} (Empirical, Em, in Table  \ref{tab:diff})), i.e. performing a linear regression between the corresponding terms in the induction equation, assuming 6$^\mathrm{th}$ order dissipation,
\begin{equation}\label{eq:nufit}
    y=\nu^F_6 s +c,
\end{equation}
where
\begin{eqnarray}
    s&=&\nabla^6 {\mathbf B}; \\ \nonumber
    y&=&\frac{\partial {\mathbf B}}{\partial t} - \nabla\times \left({\mathbf v}\times{\mathbf B}\right).
\end{eqnarray}
The regression was performed separately for all three components of ${\mathbf B}$. Unlike \cite{Popescu+etal2023}, where filtering was the only numerical stabilization mechanisms, in our case we also used hyper-diffusion in other equations. Therefore, since we are altering other variables through these terms, the theoretical results from Eq.~\ref{eq:filtcoef} can be potentially altered.  We assume that the coefficient $\nu_6^F$ is height- and time-independent. The resulting calculation is provided in Table \ref{tab:diff} and in Fig.~\ref{fig:ambi}. The values we obtained from the fit turned out to agree withing a factor of 2 with those given by  Eq.~\ref{eq:filtcoef}. Considering all the uncertainties, this can be considered a good agreement. Since $\eta_A$ is in units of m$^6$ s$^{-1}$, and $\nu_6^F$ is in units of m$^2$ s$^{-1}$, in order to compare these numbers, we have scaled $\nu_6^F$ with the grid size in each direction, i.e. divided $\nu_6^F$ by $\Delta s=\{\Delta x, \Delta y, \Delta z\}$, and we also took the average over the three spatial dimensions. Both Tab. \ref{tab:diff} and Fig.~\ref{fig:ambi} show $\nu_6^F/ds^4$. It can be seen that smaller coefficients are generally present for higher resolutions. The values obtained for the 20 km resolution cases agree well with the number given in \cite{Khomenko+etal2017} for their small-scale dynamo simulations at the same resolution. In the latter paper, erroneously a second-order diffusive filtering term was assumed, which is not the correct approximation for the filtering operation. Nevertheless, the values of the coefficients agree with a good precision. Compared to $\eta_A$ (see Fig.~\ref{fig:ambi}), one can observe that the physical ambipolar diffusion overcomes the numerical one, on average, at heights above 700-800 km. In general, the higher the resolution of the models, the lower the height where the ambipolar diffusion overcomes the numerical one, which is an expected behavior. Notice however that these average trends do not reflect in how many spatial locations the criteria $\eta_A > \nu_6^F/ds^4$ is fulfilled. We will address this question further below in Figs.~\ref{fig:la-scales} and \ref{fig:la-image}.

Having the values of $\nu_6^F$ and hyper-diffusion terms, we calculated the hydrodynamic ($Re$) and magnetic ($Rm$) Reynolds numbers, as per their definition, see Eq. 1 in \cite{Khomenko+etal2017}, including the additional filtering component of the diffusion, i.e. Eqs. \ref{eq:vdiff}, and \ref{eq:bdiff}, which was not included in the latter work. 

\begin{eqnarray}
Re&=&\frac{|\rho({\mathbf v}\nabla){\mathbf v}|}{|\nabla \cdot\boldsymbol{\tau} + \rho\nu_6^F\nabla^6{\mathbf v}|}; \\ 
Rm&=&\frac{|\nabla\times({\mathbf v}\times{\mathbf B})|}{|\nu_6^F\nabla^6{\mathbf B}|}.
\end{eqnarray}
We computed the $\nu_{j} (v_{i})$ coefficients required for $\boldsymbol{\tau}$ in Eq.\ref{eq:tau} following the same procedure as in \mancha utilizing the known values of the amplitudes of hyper-diffusion terms. It has to be noted that estimating Reynolds numbers from this kind of simulations is not straightforward and not unique since it involves the comparison of a wide range of scales. Here we attempted to use a classical definition and directly quantify the advection and diffusion terms. It order to make a more fair comparison between these terms we filtered out the scales dominated by the 6th order filtering, equivalent to 4 grid points \citep[see][for the properties of the filter]{Parchevsky+Kosovichev2007}. When computing the average values, we employed a 10\%-trimmed mean, wherein 5\% of the smallest and largest values were removed. The resulting mean values of $Re$ and $Rm$ are provided in Table \ref{tab:diff}. 

As anticipated, the Reynolds numbers scale with the resolution, explaining the higher field strengths observed in models with increased numerical resolution, as depicted in Fig.~\ref{fig:bmean} above. The Reynolds number for the Dyn$_{20}$ model differs from that in \citet{Khomenko+etal2017}, despite having the same resolution, because the equation used in the latter work to evaluate $Re$ did not include the filtering component. In this work, the correct filtering operator is used, together with an improved averaging technique that excludes extreme values.

Table \ref{tab:diff} also reflects that the $Re$ number decreases with magnetization for the same resolution, whereas $Rm$ values increase with magnetization. These trends can most certainly be attributed to the differences in scales of the flows and magnetic fields between the models.

Eventually, we proceeded to compute the artificial heating terms in the internal energy equation, as represented by Eq.~\ref{eq:ediff}, and the ambipolar term, Eq.~\ref{eq:ambidiff}. The height dependence of the horizontal and time-averaged heating terms, quantified in units of energy density per unit time, is depicted in Fig.~\ref{fig:heating-z}. We added up both the filtering  ($Q^{\rm VISC}_\mathrm{ filt}$) and hyper-diffusive components ($Q^{\rm VISC}_\mathrm{ hyp}$) of the viscous term, noting that the filtering part is approximately an order of magnitude smaller, $Q^{\rm VISC}=Q^{\rm VISC}_{\rm filt}+Q^{\rm VISC}_\mathrm{ hyp}$. It is worth noting that both $Q^{\rm MAG}_{\rm filt}$ and $Q^{\rm VISC}$ exhibit a decrease with height. The behavior of $Q^{\rm VISC}$ and $Q^{\rm MAG}_{\rm filt}$ across different models reveals an inverse scaling with resolution, in a way that there is a stronger numerical dissipation in the lower resolution models. The inverse scaling is more discernible for $Q^{\rm VISC}$, particularly for the Bz50 models. This will have repercussion into the average temperature structure, described below in Section~\ref{sect:efficiency}.

Both numerical heating terms surpass the physical ambipolar term by orders of magnitude in the sub-surface and surface layers. Only in the upper regions of the domain their values become similar. Yet, on average, the ambipolar heating term remains below the numerical counterpart, even in the highest resolution scenario. There are two prominent features: the strong scaling with resolution exhibited by $Q_\mathrm{AMB}$ (unlike the inverse scaling of the numerical terms) and its height-dependent increase, attributed to the behavior of $\eta_A$. This tendency suggests that the ambipolar heating may eventually surpass the numerical heating and become the dominant source of heating in models with even higher resolution and extending further in height.
As discussed in \citet{Khomenko+Collados2012}, the ambipolar heating would stop acting once the atmosphere is completely ionized and when the magnetic field becomes potential, leaving no more currents to dissipate. According to the models presented in \citet{Martinez-Sykora+etal2023}, the action of the ambipolar diffusion extends to heights above the transition region, approximately 3-5 Mm above the photosphere, due to the overshooting of the partially ionized material from below. Therefore we expect that the curves presented at the right panel of Fig.~\ref{fig:heating-z} would continue the growth till some heights in the low corona, where the plasma becomes eventually fully ionized. 

Despite on average $Q_\mathrm{AMB}$ stays always below its numerical counterparts, it still takes large values locally, and at certain locations it becomes the dominant contribution. Figure \ref{fig:heating-hist} illustrates the histograms of the distribution of the heating terms at 1 Mm height (left panels), together with their spatial distributions. A visual comparison of the snapshots allows to verify that the behavior in the Dyn and in the unipolar (Bz50 \& Bz200) models is rather different. The spatial distribution reveals strong viscous heating related to  acoustic shocks in the Dyn models (top second panel). At the height where the distribution is shown, the velocity variations are mostly due to waves generated in the bottom layers by the convective motions. Since the magnetic field is dynamically weak in the Dyn models, the plasma $\beta$ stays relatively large in the upper part of the domain, and the dynamics of the waves is essentially guided by hydrodynamic forces. These acoustic waves eventually steepen to shocks and their random motion and interference produce the pattern seen in at the chromospheric cuts in Fig.~\ref{fig:vzsnap-dyn} (middle panels) for velocities \citep[see also][ their Figure 16]{Khomenko+etal2018}, and the associated patterns observed at the upper panels of Fig.~\ref{fig:heating-hist} for the heating terms. 

Unlike that, in the unipolar models, the spatial distribution of $Q^{\rm VISC}$ traces the magnetic fields, because the latter tangibly affect waves and flows in the low plasma beta environment. At the heights of the cuts shown in Fig.~\ref{fig:heating-hist} the plasma $\beta$ is well below one. Therefore, fast acoustic waves generated by convection at lower heights are converted into the fast/slow magneto-acoustic waves, the latter propagating field-aligned \citep[identical to the behavior reported in][see their Figure 17]{Khomenko+etal2018}. This changes the visual aspect of the velocity snapshots in Fig.~\ref{fig:vzsnap-uni} (middle panels), and the corresponding distributions of the heating terms. The field-aligned propagation produces a strong shear in velocity and therefore the strongest $Q^{\rm VISC}$ is unipolar models corresponds to the locations of slow field-aligned magneto-acoustic wave fronts. The numerical magnetic heating $Q^{\rm MAG}_{\rm filt}$, follows the locations of the strongest currents in all the models (third column of panels). 

The ambipolar heating, $Q_{\rm AMB}$ (last column) follows similar patters as described above for each of the models, but additionally it is enhanced around the cool and ratified locations corresponding to propagating wave fronts. This behavior is particularly evident in the Dyn case. In this case, it can be observed that, frequently, at locations with the lowest $Q^{\rm MAG}_{\rm filt}$ (black areas), $Q_{\rm AMB}$ is enhanced (diffuse gray areas). The ambipolar heating in these locations overcomes the viscous one as well. In the unipolar cases, filamentary structures with enhanced $Q_{\rm AMB}$ also fill the darkest areas present in $Q^{\rm MAG}_{\rm filt}$, with amplitudes comparable or even larger than the viscous heating. 

The histograms of the distribution of the heating terms, shown in Fig.~\ref{fig:heating-hist} (left), verify that $Q^{\rm VISC}$ provides the largest heating in all the studied cases. The vertical green dotted lines, which mark the maxima of the $Q^{\rm VISC}$ distributions are always to the right from the blue and magenta lines for $Q^{\rm MAG}_{\rm filt}$ and $Q_{\rm AMB}$, correspondingly. A similar dominance of $Q^{\rm VISC}$ was observed in the turbulent flow created in the non-linear two-fluid simulations of the Rayleigh-Taylor instability in solar prominences \citep{Popescu+etal2023, Lukin+etal2024}, however there the origin of the viscosity was physical and not numerical. 

The histograms in Fig.~\ref{fig:heating-hist} depict that the relative importance of the $Q^{\rm MAG}_{\rm filt}$ and $Q_\mathrm{AMB}$ depends on the resolution and the magnetization of the model. Both parameters have similar distributions in the Dyn$_5$ and Bz50$_{10}$ runs (the vertical blue and magenta lines almost coincide). Thus, the importance of both processes is similar at chromospheric heights. In the Bz200$_{20}$ case, $Q_\mathrm{AMB}$ becomes, on average, larger than  $Q^{\rm MAG}_{\rm filt}$ (the vertical magenta line is to the right from the vertical blue line). Therefore, we conclude that $Q_{\rm AMB}$ introduces a relevant contribution into the heating of the upper layers of our models, comparable to or even larger than $Q^{\rm MAG}_{\rm filt}$.

\begin{figure}
\centering
\includegraphics[keepaspectratio,width=8cm]{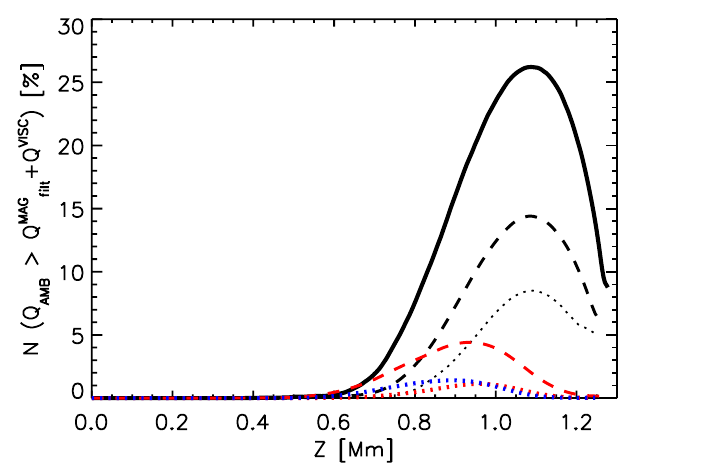}
\includegraphics[keepaspectratio,width=8cm]{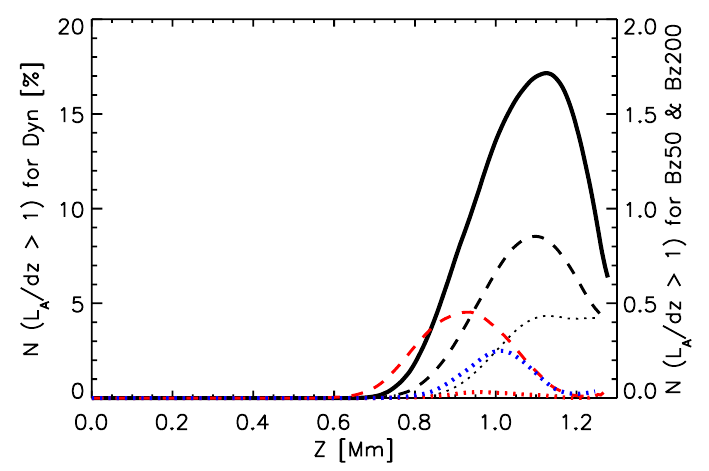}
\includegraphics[keepaspectratio,width=8cm]{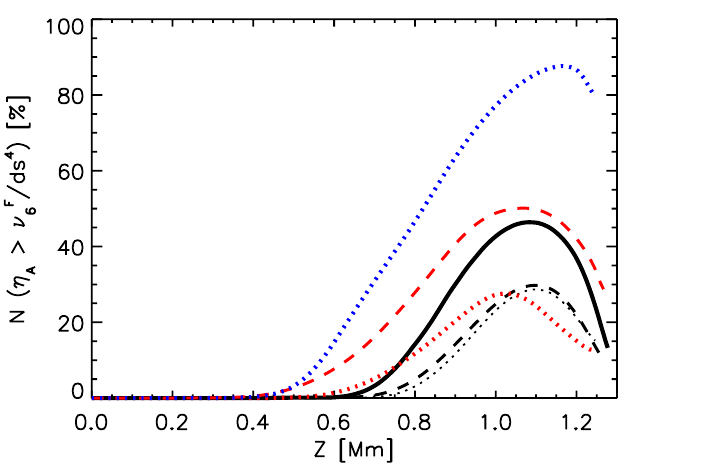}
\caption{\footnotesize  Top: Percentage of the spatial locations where the condition $Q_\mathrm{AMB} > Q^{\rm VISC}_{\rm filt}+Q^{\rm VISC}_\mathrm{ hyp} +Q^{\rm MAG}_{\rm filt}$ is fulfilled for the different models, shown as a function of height. Black lines are for Dyn model, red for Bz50 model and blue for Bz200 model. Solid thick line is for 5 km resolution, dashed medium-thick line is for 10 km resolution, dotted thin line is for 20 km resolution. Middle: same for $L_A>dz$.  Bottom: same for $\eta_A > \nu_6^F/ds^4$.}
\label{fig:la-scales}
\end{figure}

\begin{figure*}
\centering
\includegraphics[keepaspectratio,width=12.5cm]{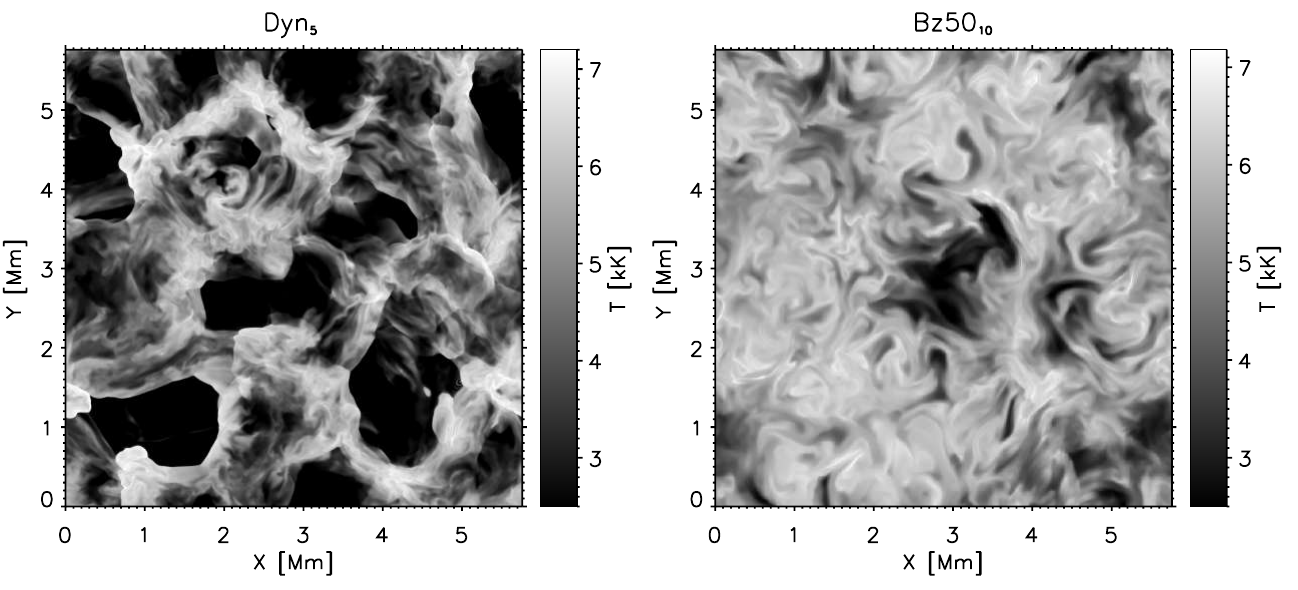}
\includegraphics[keepaspectratio,width=12.5cm]{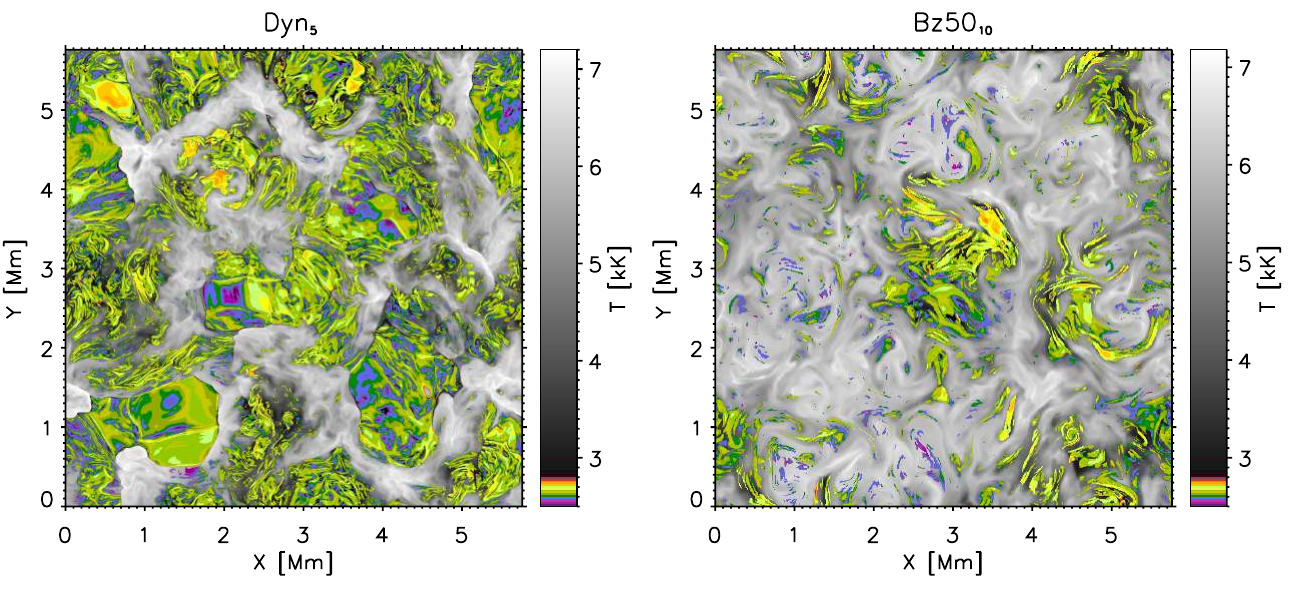}
\includegraphics[keepaspectratio,width=12.5cm]{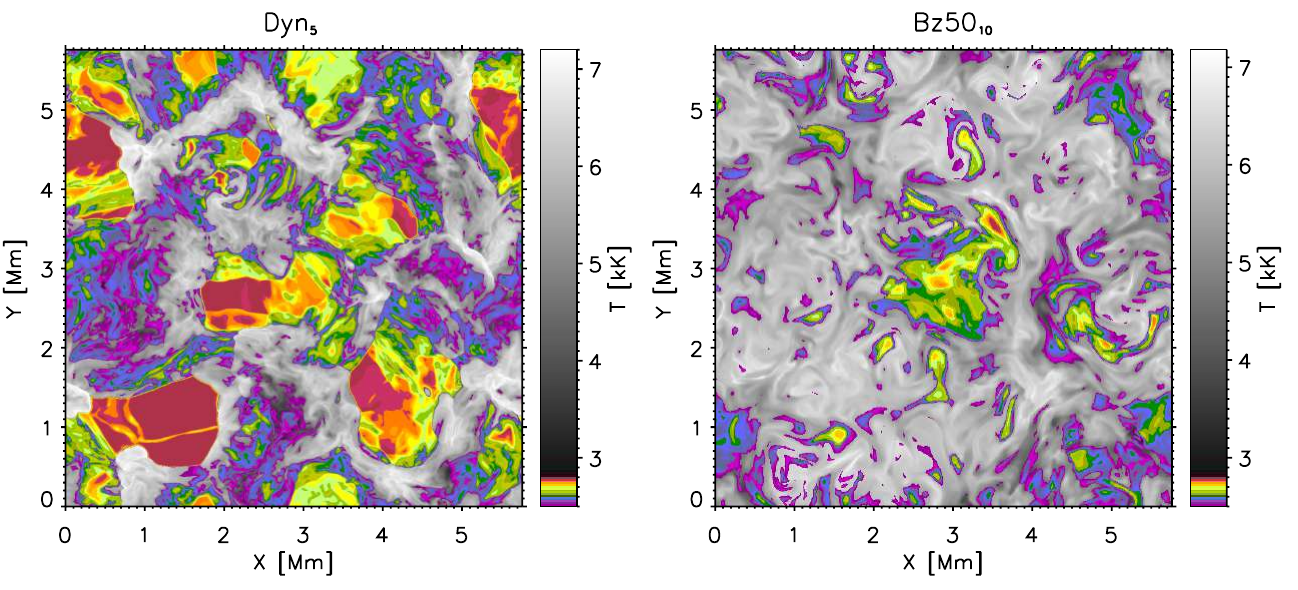}
\includegraphics[keepaspectratio,width=12.5cm]{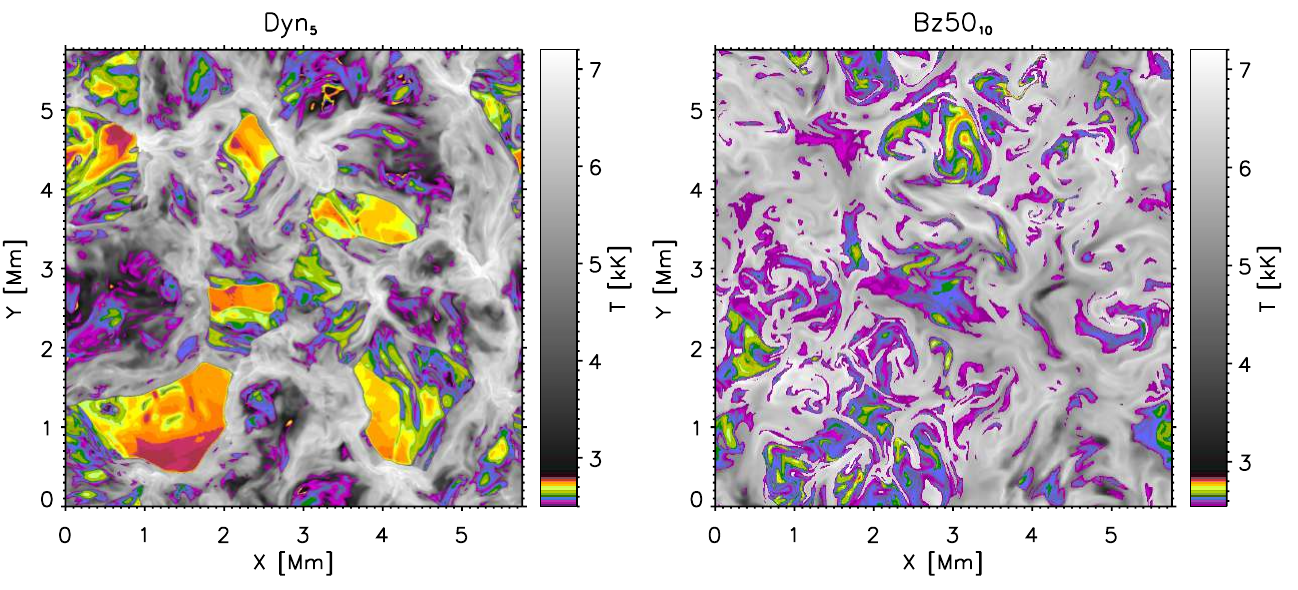}
\caption{\footnotesize First row: Snapshot of temperature at z=1 Mm in the Dyn$_5$ (left) and in the Bz50$_{10}$ models (right) for comparison purposes.  Second row: Locations with the largest  $Q_{\rm AMB}$ from $Q_{\rm AMB}=10^{-6}$ J m$^{-3}$ s$^{-1}$ (violet color) to $Q_{\rm AMB}=10^{-1}$ J m$^{-3}$ s$^{-1}$ (vermilion color), plotted over the temperature image (in gray) only at locations with the condition $Q_{\rm AMB}>0.1*Q_{\rm scheme}$ fulfilled. Third row: same, except that colored areas indicate locations with the largest $L_A/dz$ in logarithmic scale, in a range from 0.01 (violet color) to 10 (vermilion color) at locations with the condition $L_A/dz > 0.01$ fulfilled. Forth row: same, except that colored areas indicate locations with $\eta_A > 2\nu_6^F$ with $\eta_A$ in the range from $10^6$ to $10^9$ m$^2$s$^{-1}$.}
\label{fig:la-image}
\end{figure*}

\section{Ambipolar heating scale}
\label{sect:efficiency}

An alternative way of verifying how well our simulations resolve the physical scales of the ambipolar heating is to compare the characteristic scales of this effect with the numerical grid size. Following \citet{Khomenko+etal2014}, we define the ambipolar scale, $L_A$, as the ratio between the ambipolar coefficient (Eq.~\ref{eq:etaa}), and the characteristic speed of propagation of magnetic perturbations, i.e. the Alfv\'en speed, $v_A=B/\sqrt{\rho\mu_0}$,
\begin{equation}\label{eq:la}
L_A=\frac{\eta_A}{v_A}.
\end{equation}

For all the simulations, we have computed the number of spatial locations that satisfy the criteria of either resolving the ambipolar scale (i.e., $L_A > dz$), or having the physical ambipolar heating exceed the numerical one (i.e. $Q_\mathrm{AMB} > Q_{\rm scheme}$, where $Q_{\rm scheme}=Q^{\rm VISC}_{\rm filt}+Q^{\rm VISC}_\mathrm{ hyp} +Q^{\rm MAG}_{\rm filt}$).  For comparison, we also computed the percentage of locations where the ambipolar diffusion is above the numerical one, $\eta_A > \nu_6^F/ds^4$. The results of this computation are displayed in Figure \ref{fig:la-scales},  which shows the percentage of points in the horizontal plane that satisfy a given criterion as a function of height.

It can be seen that $L_A$ and $Q_\mathrm{AMB}$ criteria produce qualitatively similar results. The fraction of resolved points is negligible below approximately 0.6 Mm and shows a maximum between 0.9 and 1.1 Mm, depending on the model. The increase in the fraction of resolved points with height is due to the increase in the ambipolar diffusion coefficient. The shape of the curves above the maximum is most likely a consequence of the upper boundary condition. 

We observe an expected scaling with resolution: the number of resolved points approximately doubles between consecutive resolutions for the $L_A$ and $Q_\mathrm{AMB}$ criteria. For both of these criteria, there are significantly more resolved points in the Dyn models compared to the unipolar models. In the Dyn models, we resolve the ambipolar effect in a maximum of 17\% of points based on the $L_A$ criterion, and 27\% of points based on the $Q_\mathrm{AMB}$ criterion. These numbers are significantly smaller in the unipolar models, where the number of resolved points does not exceed a few percent. As can be seen by comparing the vertical axes of the upper and middle panels of Fig.~\ref{fig:la-scales}, applying the $Q_\mathrm{AMB}$ criterion produces a larger fraction of resolved points in the unipolar models (notice that the left panel has a second vertical axis for the curves from Bz50 and Bz200).

The disparity between the models results from the lower values of $L_A$ in the unipolar models. If we consider the main dependencies for $L_A$ in Eq.~\ref{eq:la}, taking into account the definition of the ambipolar diffusion coefficient, Eq.~\ref{eq:etaa}, it appears that $L_A \sim B/P_e\sqrt{T/\rho}$, where $P_e$ is the electron pressure and $T$ is temperature. The value of $\sqrt{T/\rho}$ only weakly depends on the model, while $B$ is on average larger in the unipolar models, see Fig.~\ref{fig:bmean}. Nevertheless, the temperature is on average also larger at the chromospheric heights in the unipolar models, as it will be shown below in Fig.~\ref{fig:temp_height_eta}. These larger temperatures produce larger ionization fractions and larger values of the electron pressure, $P_e$. The latter, through the inverse dependence, overcompensates the increase of $L_A$ due to large magnetic fields, and causes lower values of $L_A$ in the unipolar models. 

The $\eta_A$-base criterion (bottom panel of Figure \ref{fig:la-scales}) leads to somewhat different conclusions. Firstly, there is a significant number of points with $\eta_A > \nu_6^F/ds^4$ deeper down into the atmosphere, compared to the other two criteria. Secondly, the ordering of the curves for the models with different magnetization is reversed. Now the models with larger magnetization show a much larger fraction of points fulfilling the criterion. Finally, the amount of points fulfilling $\eta_A > \nu_6^F/ds^4$ is significantly larger than for the $Q_\mathrm{AMB}$ and $L_A$ criteria. These differences are produced by the fact that the $\eta_A$-base criterion is much less restrictive and only takes into account the actual value of the ambipolar coefficient. The $L_A$ criterion considers additionally the speed of propagation of the signal between the numerical grid points, leading to a reduction of the scale in the strongly magnetized unipolar models. Nevertheless, it needs to be kept on mind that $L_A$ estimates are only an approximation. We consider the $Q_\mathrm{AMB}$ criterion being the most complete of the three since it includes the comparison between the ambipolar heating and all other types of heating, including the viscous one. 

The main conclusion from Figure~\ref{fig:la-scales} is that at our best horizontal resolution of 5 km, we numerically resolve the ambipolar diffusion in a maximum of 27\% of points based on the more complete $Q_\mathrm{AMB}$ criterion. In models extending higher up and less affected by the vertical magnetic field at the upper boundary, it is expected that the percentage of resolved points would continue to increase with height due to the drop in density, leading to larger values of  $\eta_A$. However, this increase would be strongly conditioned by the specific magnetic field configuration in a given model.

Figure \ref{fig:la-image} shows the spatial distribution of locations with large $Q_{\rm AMB}$  (second row of panels), large $L_A$ (third row of panels), and large $\eta_A$  (forth row of panels) plotted over an image of temperature at chromospheric heights where the abundance of the resolved points, according to each of the criteria, is at its maximum, in two representative cases, Dyn$_5$ and Bz50$_{10}$. Visibly, there are fewer resolved points in the unipolar model. It is observed that $L_A$ and $Q_\mathrm{AMB}$ criteria give similar results. The comparison of these locations with the temperature image reveals that most of the resolved points correspond to low-temperature regions in the chromosphere. This is not surprising, given the fact that in these cool regions the neutral fraction is the largest and the density is the smallest, producing maximum values of $\eta_A$. 

Nevertheless, the bottom panel of Fig.~\ref{fig:la-image} shows that the locations of the points with the largest $Q_{\rm AMB}$, $L_A$, and $\eta_A$ are not identical, especially in the unipolar model. Specifically, $\eta_A$ values are indeed higher in the coldest regions. The correlation between all three criteria is strong in the dynamo model because the magnetic field is weak, and the distribution of $\eta_A$ is primarily influenced by the thermodynamic properties of the chromospheric plasma. In the unipolar models, the locations with large $\eta_A$ exhibit somewhat different distributions. While $\eta_A$ remains large in the cold areas, it is also visibly enhanced in the strongly magnetized, yet not as cold, regions.

It is possible that our conclusions are affected by considering equilibrium ionization values. Taking into account time-dependent ionization is known to produce over-ionized areas at the wakes of acoustic shocks, or nearly-adiabatically expanded material in the chromosphere \citep{Leenaarts2007, Nobrega-Siverio+etal2020}. Nevertheless, it has to be taken into account that, alongside with the ionization fraction, density drop with height also causes a strong effect on $\eta_A$, so the fraction of the resolved points in the simulations can be expected to increase with height, beyond the cool areas.  

\section{Scaling of heating with resolution and magnetization}
\label{sect:scaling}

\begin{figure*}
\centering
\includegraphics[keepaspectratio,width=14cm]{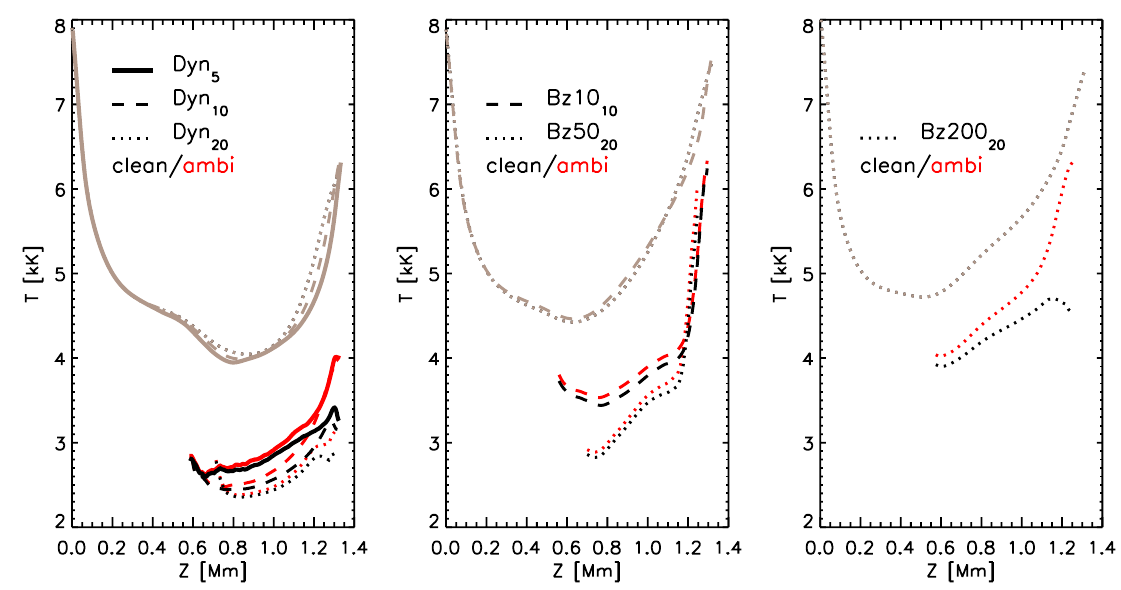}
\caption{\footnotesize Height dependence of the average temperature (light brown curves extending over all height range) together with the average temperature at locations with $Q_{\rm AMB}> Q^{\rm VISC}_{\rm filt}+Q^{\rm VISC}_\mathrm{ hyp} +Q^{\rm MAG}_{\rm filt}$  in the ``clean'' runs (black) and in the ``ambi'' runs (red). Panels from left to right are for the Dyn,  Bz50, and Bz200 models. The averaging is performed for the times from 200 sec till 1200 sec for Dyn$_{20}$, Dyn$_{10}$, and Bz50$_{20}$, and from 200 sec till the maximum available time in the rest of the cases. }
\label{fig:temp_height_eta}
\end{figure*}

\begin{figure*}
\centering
\includegraphics[keepaspectratio,width=14cm]{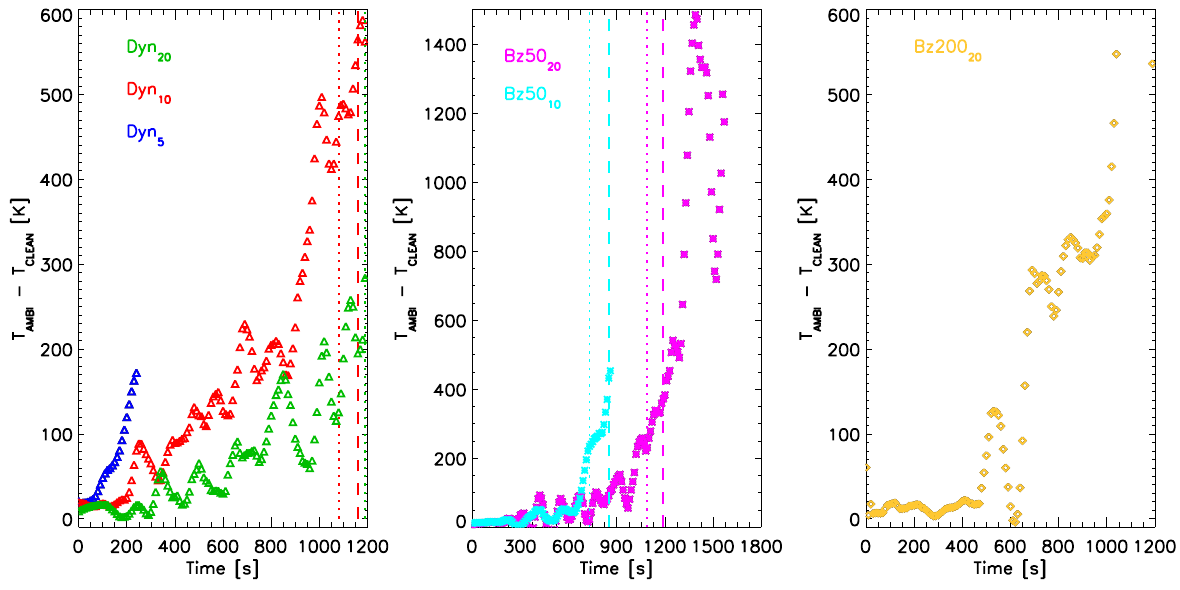}
\includegraphics[keepaspectratio,width=14cm]{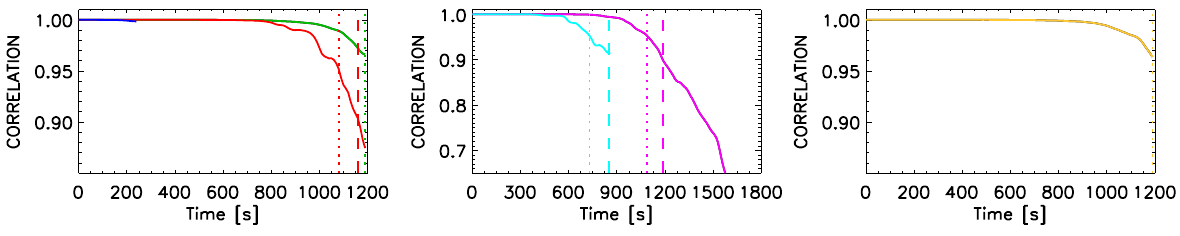}
\caption{\footnotesize Top panels: temperature difference between ``ambi'' and ``clean'' runs as a function of time. The temperature difference is averaged over heights where the percentage of points satisfying the criteria $Q_{\rm AMB}>Q^{\rm VISC}_{\rm filt}+Q^{\rm VISC}_\mathrm{ hyp} +Q^{\rm MAG}_{\rm filt}$ is at least 20\% of the maximum number of points satisfying this criteria for a given model (as per Fig.~\ref{fig:la-scales}).  The curves are smoothed over 5 data points to decrease the scatter and appreciate the trends. Left: Dyn runs, middle: Bz50 runs, right: Bz200 run.  Bottom panels: correlation coefficient between temperature in the ``ambi'' and ``clean'' runs as a function of time. Vertical lines mark times where the correlation drops to 95\% (dotted lines) and 90\% (dashed lines), where appropriate. Note the difference in the axes range between the middle panels compared to the left and right panels.}
\label{fig:dtemp_time_qamb}
\end{figure*}

\begin{figure*}
\centering
\includegraphics[keepaspectratio,width=16cm]{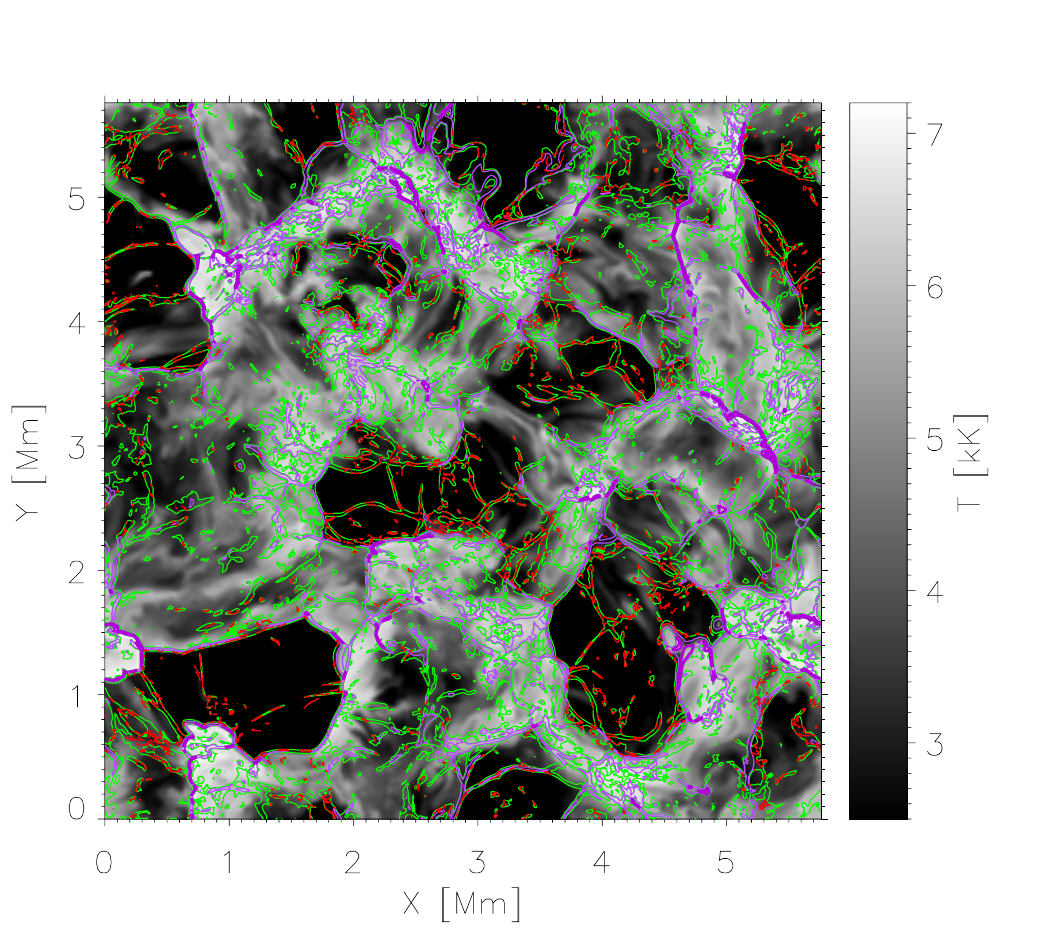}
\caption{\footnotesize Spatial variation of temperature at 1 Mm height in the Dyn$_5$ model (gray color), together with iso-contours of compressional heating $Q_{\rm COMP}$=0.001 (green), 0.1 (light violet), 1 (dark violet) J m$^{-3}$ s$^{-1}$, where a larger number means a larger compression. Red colors show the iso-contours of a mask where two conditions together are fulfilled: $Q_{\rm AMB} > Q_{\rm scheme}$ and $Q_{\rm COMP} >0$. }
\label{fig:te_qcomp}
\end{figure*}

Having identified the locations where the physical ambipolar heating is resolved, we computed the temperature differences between the runs with ambipolar diffusion and the "clean" runs.
To facilitate the comparison between the runs, we have created masks based on either of the three criteria used above (e.g. $Q_{\rm AMB} > Q_{\rm scheme}$), and then performed averages of the temperature for the points inside these masks. The masks were computed using the variables from the ``clean'' runs, and then applied to their ``ambi'' counterparts to assure that we consider exactly the same points in space. During the first 1200 sec of time the granulation patterns in the ``clean'' and ''ambi'' simulations are similar enough, so that the masks based on either of them produce very similar results. Nevertheless, to have a cleaner experiment we base the results shown in Figs.~\ref{fig:temp_height_eta} and \ref{fig:dtemp_time_qamb} below on the masks computed over the ``clean'' runs. This is done because ambipolar diffusion is not acting in the "clean" runs and the heating produced by it does not "contaminate" the definition of the locations where the criterium is satisfied.

In Figure \ref{fig:temp_height_eta}, we show the time-averaged temperatures at locations with ambipolar heating exceeding the numerical heating in the "ambi" runs (red curves). For comparison, we also averaged the temperatures at exactly the same locations in the "clean" runs, shown in black.  We performed the time averaging from 20 seconds of simulation time until the maximum of 1200 sec, when available, as the differences increase over time. Note that, due to the fewer snapshots available for the Dyn$_5$ case, the averaging covers a much shorter period compared to the other cases. The curves are only shown for heights above approximately 0.6 Mm, where a significant number of points meet the selection criteria. These time-averaged curves are compared to the time- and horizontally-averaged temperatures across the entire domain in each simulation (light brown curves). The domain-averaged and the selected location-averaged curves can be easily distinguished from each other, as the selection criteria pick the locations with the lowest temperatures.

A few conclusions can be extracted from Figure \ref{fig:temp_height_eta}. Firstly, one can observe that the time-average difference between the "clean" and the "ambi" runs scales with the resolution. Even if a much shorter time span is available in the Dyn$_5$ case, the temperature increase is more significant than in the Dyn$_{10}$ and Dyn$_{20}$ cases (left panel). The differences also grow with height, reflecting the growth of the ambipolar heating. A similar behavior is observed in the Bz50 case, where the two resolutions can be compared. Secondly, we observe that, over most of the height range, the locations with large $Q_{\rm AMB}$ correspond to progressively larger temperatures for progressively finer resolutions. Notice however, that on average the higher resolution models are slightly cooler (light brown curves). The latter is a consequence of the inverse scaling with resolution of $Q_{\rm VISC}$ and $Q^{\rm MAG}_{\rm filt}$, see Fig.~\ref{fig:heating-z}. For the explanation of the former, we refer to Figure \ref{fig:te_qcomp}. This figure shows a snapshot of the temperature at 1 Mm height in the Dyn$_5$ model together with the contours of compressional heating, $Q_{\rm COMP}=-p(\nabla\cdot{\mathbf v})$, where $p$ is the gas pressure. Notice that $Q_{\rm COMP}$ can take both positive and negative values. Positive values correspond to compression, typically obtained at the shock fronts at chromospheric heights. The locations of maximum compression are right at the borders between the hot and cold chromospheric material, but the compression also extends into the cool areas right in front of the shocks. Red contours in Fig.~\ref{fig:te_qcomp} denote the locations where both conditions are fulfilled, $Q_{\rm AMB} > Q_{\rm scheme}$ and $Q_{\rm comp} >0$, corresponding to compression. It can be seen that, while they do not coincide with locations of maximum compression (violet contours), they do frequently coincide with the green contours delineating mild compression inside the cool areas. The compressional heating is rather strong, the value at the green contours, $Q_{\rm comp} =0.01$ J m$^{-3}$ s$^{-1}$ is above the typical values of $Q^{\rm VISC}$ and $Q^{\rm MAG}_{\rm filt}$ (Fig.~\ref{fig:heating-z}, \ref{fig:heating-hist}). This leads us to conclude that the locations selected by the $Q_{\rm AMB} > Q_{\rm scheme}$ criterion are additionally heated by compression both in the ``clean'' and ``ambi'' models. The strength of the compressional heating scales with the resolution due to the sharpening of the gradients, therefore, models with higher resolution undergo larger compressional heating, which produce, on average, larger temperatures at the locations selected by the $Q_{\rm AMB} > Q_{\rm scheme}$ criterion, seen in Fig.~\ref{fig:temp_height_eta} (black and res curves).

Fig.~\ref{fig:temp_height_eta} additionally shows that the heating happens at progressively hotter areas in models with progressively larger magnetization. The unipolar models do not produce such low temperatures as the dynamo ones, since they are heated by a larger viscous and Joule dissipation (compare red, blue and black lines in Fig.~\ref{fig:heating-z}). Finally, we find that the temperature increase is the largest in the Bz200 case. This is different from the findings reported in \cite{Khomenko+etal2018}, where the largest heating was found to happen in the Dyn model. Nevertheless in the latter work only a unipolar model of 10 G was studied, while here we are checking larger magnetizations. The explanation provided in \cite{Khomenko+etal2018} was that the heating is more efficient in the Dyn cases since it happens in the deeper layers where the additional energy is spent into the temperature increase and not into hydrogen ionization. While this conclusion still holds, we now observe a significantly larger ambipolar heating in the Bz200 model, even if we resolve only a small portion of the domain. These curves slightly change if the selection of points is based on the $L_A$ rather than on $Q_{\rm AMB}$, but the main conclusions remain.

\begin{figure*}[ht]
\centering
\includegraphics[keepaspectratio,width=18cm]{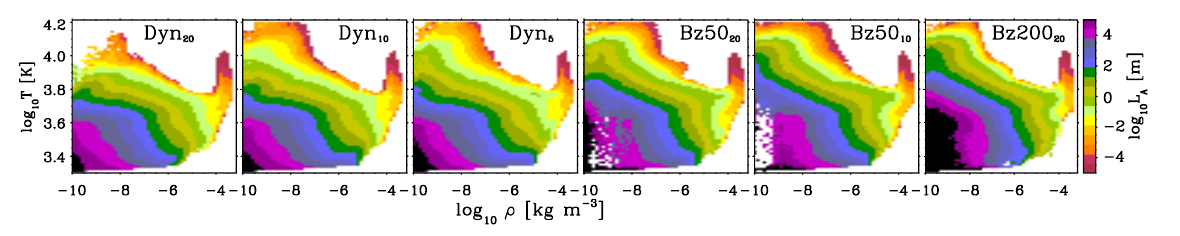}
\caption{\footnotesize Ambipolar scale, $L_A$ as a function of density and temperature in all the models (indicated in each panel).}
\label{fig:la_te_dens}
\end{figure*}

Additionally, in the upper panel of Figure \ref{fig:dtemp_time_qamb} we plot in symbols the temperature difference, $T_{\rm ambi} - T_{\rm clean}$ as a function of time. The bottom panel of this figure shows the correlation between the temperatures in the "ambi" and "clean" runs, marking the location where the correlation drops to 95\% and 90\%, where appropriate. We observe that the temperature difference between the "ambi" and "clean" runs in the resolved locations increases in time in the interval from 0 till 1200 s, or slightly longer as in the case of the Bz50$_{20}$ run, which was continued in time to check our hypothesis (see the discussion below). It can be observed that the values are always positive. During the 1200 sec of time the temperature in all the models increases by a maximum of 600 K. The increase is the fastest in the Bz200 case. The curves are steeper in progressively larger resolutions. Similar conclusion is reached when the points are selected based on the $L_A$. The particular shape of the curves depends to some extent on the criteria used for the averaging and on the chosen thresholds. Despite only a short series is available in the Dyn$_5$ case, it can be seen that the efficiency of the heating in this model is the highest, and at similar time moments, higher temperature differences are reached. Due to the intrinsic nature of granulation, it is not possible to trace this behavior much longer in time, since the granulation pattern varies and the snapshots cannot be compared one to one. The structure's boundaries change shape and position in the "ambi" and "clean" runs and the averages become affected by these changes, which might lead to a misinterpretation of the results, since the temperature averages at locations selected by a given criteria in the pair of simulations would not correspond to the same structures. This behavior reflects itself in the drop of the correlation (bottom panel). Nevertheless, the correlation keeps above 90\% or even 95\% in all of the cases during 1200 s. We regard this as proof that we are comparing fundamentally the same structures, given that ambipolar diffusion introduces a significant perturbation at the selected locations.

We checked by how much longer we could potentially compare the "ambi" and "clean" runs by continuing the Bz50$_{20}$ simulations for another 300 sec in time. The result is shown at the middle panel of Figure \ref{fig:dtemp_time_qamb}. It can be seen that the curve reaches a maximum of about 1400 K at 1400 sec and then drops and shows some oscillation. At the same time, the correlation decreases to 60\% after 1500 s. This verifies the expected behavior and the impossibility of doing a meaningful comparison between the simulations for much longer time.  Nevertheless the conclusions are clear: even in our best resolution case we do not reach the saturation of the ambipolar heating, it produces larger temperatures as time goes by, and the amount of temperature increase in the Dyn case almost scales with the resolution. 

\begin{figure}
\centering
\includegraphics[keepaspectratio,width=9cm]{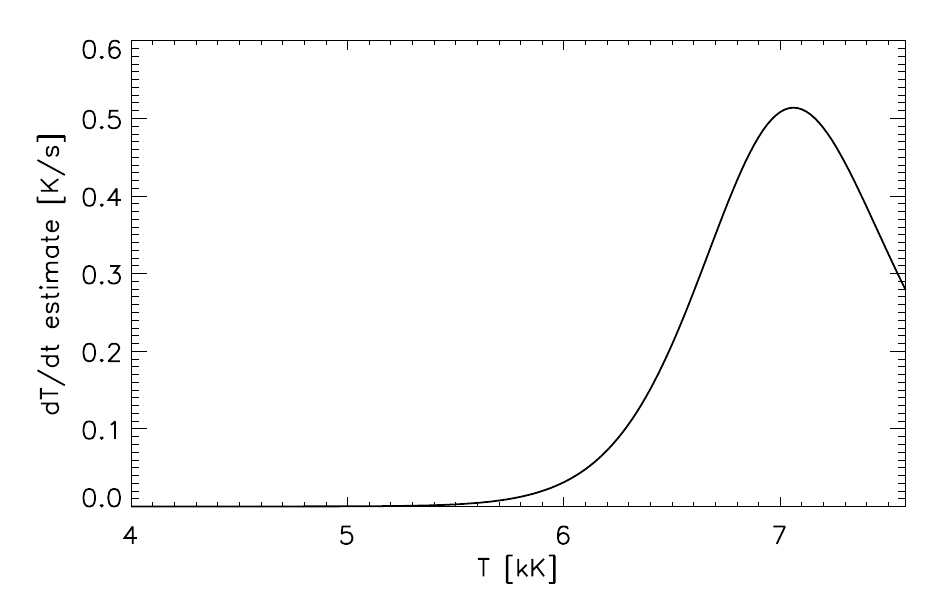}
\caption{\footnotesize  Approximate temperature variation rate due to $Q_{\rm AMB}$, computed following Eq.~\ref{eq:tevol}, as a function of temperature.}
\label{fig:trate}
\end{figure}

Another feature observed in Figure \ref{fig:dtemp_time_qamb} is the intermittent behavior in time and apparently non-linear trend of the temperature curves, especially pronounced in the case of the unipolar runs (middle and right panels). For small perturbations, it was previously observed that $Q_{\rm AMB}$ can be considered temperature-independent leading to a linear temperature increase over time, see for example Figure 11 in \citet{Popescu+etal2019}. Nevertheless, $Q_{\rm AMB}$ is a function of temperature through the temperature dependence of the neutral fraction $\xi_n$ and the collisional parameter $\alpha_n$, and for larger perturbations, this can lead to loss of linearity in $Q_{\rm AMB}$.
To illustrate this point, we consider a highly simplified case of pure hydrogen plasma of a given density ($\rho$) and electron concentration ($n_e$). In that case, the internal energy evolution equation (Eq.~\ref{eq:ambidiff}) can be transformed into the equation of the temperature evolution:
\begin{equation} \label{eq:tevol}
\frac{\partial T}{\partial t} + (...) 
\approx \frac{(\gamma-1)m_H}{\rho k_B}\frac{\xi_n^2B^2}{\alpha_n}J_{\perp}^2,
\end{equation}
with $\xi_n=n_n/(n_n+n_e)$, $k_B$ being the Boltzmann constant, $m_H$ proton mass, and $\gamma=5/3$ specific heat ratio. By neglecting the contribution of electrons and of other chemical elements apart from hydrogen, the collisional parameter $\alpha_n$ is approximately computed as,
\begin{equation}
\alpha_n=m_H n_e n_n\sqrt{\frac{16k_B T}{\pi m_H}}\Sigma_{in},
\end{equation}
where $\Sigma_{in}=5\times10^{-19}$ m$^2$ is the collisional cross section \citep{Braginskii1965}.
The neutral number density is estimated by the Saha equation,
\begin{equation}
n_n=n_e^2\left( \frac{h}{\sqrt{2\pi k_B m_e T }} \right)^3\exp{\left(\frac{\chi_{\rm ion}}{k_B T} \right)},
\end{equation}
with $m_e$ being electron mass, and $\chi_{\rm ion}=13.6$ eV is hydrogen ionization potential from the ground level.

From these simplified equations, we demonstrate that the temperature variation rate becomes a function that involves a power function of T multiplied by the exponential term $\exp{\left(\frac{\chi_{\rm ion}}{k_B T} \right)}$. Due to these dependencies, the rate is an increasing function of $T$ at lower temperatures, while it starts to decrease with a further increase of $T$ as hydrogen approaches its full ionization.

The rate computed using Eq.~\ref{eq:tevol} is presented in Figure~\ref{fig:trate}. For this computation, we fixed the density, electron concentration, magnetic field, and currents to the spatially and temporally averaged values of the Bz200$_{\rm 20}$ model at a height of 1 Mm. We then varied the temperature within a range from -1500 to 2000 K relative to the temperature at 1 Mm. We observe that the curve exhibits the expected behavior. Within the temperature range where ambipolar diffusion is acting in our models, the rate is an increasing function of temperature. Moreover, even under such rough approximations, the computed values are consistent in order of magnitude with the temperature increase over 1200 seconds observed in our simulations, as shown in Fig. \ref{fig:dtemp_time_qamb}. Therefore, this approximate calculation can be considered as an additional proof that the apparent non-linear trend of temperature curves in Fig. \ref{fig:dtemp_time_qamb} is mostly the result of the intrinsically non-linear behavior of $Q_{\rm AMB}$, rather than a loss of correlation between the compared snapshots of the ``ambi'' and ``clean'' runs due to advection.

\section{Discussion and conclusions}
\label{sect:conclusions}

In this paper, we have performed a convergence study of the ambipolar effect in realistic MHD simulations. 
We obtained that at progressively higher resolutions there is progressively more magnetic energy available for dissipation because magnetic fields are, on average, more intense. Other than that, models at different resolutions are statistically equivalent. We have performed a very careful evaluation of the numerical dissipation effects in our models. This was done by replicating the numerical viscosity and magnetic diffusivity as internally calculated in \mancha code, in which we have a full control over the numerical effects. We have identified that the physical ambipolar coefficient becomes, on average, larger than the numerical one from the lower chromosphere upwards. The precise height depends on the magnetization of the model and on the resolution. 
This analysis allows us to firmly claim that we reach resolutions where the physical ambipolar mechanism becomes the dominant one in the simulations and it is fully numerically resolved in a significant fraction of the simulated volume. By performing a one to one comparison of the snapshots with/without ambipolar diffusion, we established that the temperature at the locations where the latter is resolved increases in time. This increase shows no saturation during the 1200 sec of the simulation time, and, more importantly, it scales with the resolution also without saturation. Further continuation of the Bz50$_{20}$ run in time shows a saturation after about 1400 sec, which is attributed to the change of the granulation pattern leading to the impossibility to trace this effect further in time using this kind of simulations. Our analysis establishes that at a horizontal numerical resolution of 5 km we already resolve the ambipolar diffusion in about a quarter of the volume in the simulations of small scale solar dynamo. Since the contribution of this process increases with height due to the density drop, models with a higher upper boundary at the same resolution are more adequate to better resolve this action. The situation is less optimal in scenarios with a larger magnetization, since the ambipolar scale, $L_A$, is yet very small. Even so, one can expect to resolve ambipolar diffusion in a large fraction of the volume in the upper atmosphere and close to the transition region, where some amount of neutrals might be present.

In a turbulent flow with many nonlinear dissipative effects acting simultaneously, the effect that acts on the largest scale will be the one that provides the most significant transfer of mechanical or magnetic energy into thermal energy. This is because the amount of available energy usually scales as a power law. In the weakly ionized lower solar atmosphere, the effect acting on the largest scales is ambipolar diffusion. Our analysis reveals that the ambipolar diffusion  efficiently heats the initially coolest areas of the solar volume. In a series of works \citet{Brandenburg2011a, Brandenburg2011b, Brandenburg2014, Brandenburg+Rempel2019} discussed the dominance of the energy dissipation mechanisms in different types of dynamos as a function of Prandtl number. \citet{Brandenburg+Rempel2019} have shown that in small-scale dynamos at large magnetic Prandtl numbers, the magnetic Lorentz force does work on the flow at small scales, and the energy of fluid motions is then dissipated through viscosity. According to \citet{Brandenburg2014}, the ratio of the kinetic to magnetic energy dissipation always increases with the Prandtl number. Despite these works did not deal with the ambipolar diffusion, they clearly shown that the energy dissipation is dominated by the process with the larger length scale, i.e. the largest diffusivity. These conclusions have a clear parallelism with our work. Similar to \citet{Brandenburg2011a, Brandenburg2011b} we obtain that for large Prandtl numbers (i.e. when viscosity dominates over magnetic diffusion, which is the case in this work) most of the energy is dissipated through viscosity. This is the consequence of how our simulations were set up, i.e. by setting to zero the Ohmic hyperdiffusivity while keeping the hyperviscosity. Despite the ambipolar diffusion dominates locally, the hyperviscosity dominates on average. This viscous dissipation directly competes with ambipolar diffusion. According to \citet{Brandenburg+Rempel2019}, in this regime the presence of viscosity causes an energy transfer via the Lorentz-force from magnetic to kinetic energy on the viscous scale, which diminishes the amount of magnetic energy that can be dissipated on smaller scales through the Joule dissipation (related to the ambipolar diffusion in our case). This conclusion is important in establishing of how much of the energy can be dissipated through the ambipolar diffusion eventually, in different regimes of the Prandtl number. More works need to be done in this direction in the future.

In our simulations used in previous works, we did not observe a strong heating effect due to an insufficient resolution. When examining the average quantities, the mean temperature increase was found to be highest in the small-scale dynamo simulations compared to the unipolar field simulations \citep{Khomenko+etal2018, Gonzalez-Morales+etal2020, Khomenko+etal2021}. After the analysis performed in the current work, we have gained a full understanding of this result. In the SSD models, ambipolar diffusion begins heating in deeper areas, where this additional energy is converted more into temperature increase and less into hydrogen ionization. Additionally, the unipolar model with a 10 G mean field did not have sufficient ambipolar diffusion, nor was it numerically resolved in a significant portion of the domain at 20 km horizontal resolution. The current analysis shows that the ambipolar heating indeed becomes much larger in the unipolar models with stronger magnetization than in the SSD ones. Unfortunately, in the current work, we did not run simulations at higher resolution for the 50 G and 200 G cases due to their high numerical costs. However, even with the simulations we have, we observe that higher magnetizations produce steeper heating curves over time and larger magnitudes of temperature increase at the resolved points compared to the SSD runs.

It has been observed in our models, as well as in other similar models \citep{Nobrega-Siverio+etal2020, Martinez-Sykora+etal2017, Martinez-Sykora+etal2020}, that ambipolar heating is more efficient in the coolest areas due to lower densities and larger neutral fractions. These cool areas can appear, for example, in the wakes of shock waves or inside nearly adiabatically expanding cool bubbles during flux emergence processes. It has also been shown that the non-equilibrium ionization can significantly change the amount of ambipolar heating. \citet{Martinez-Sykora+etal2020, Nobrega-Siverio+etal2020} included ambipolar diffusion and non-equilibrium ionization of hydrogen and helium in their 2.5D simulations of flux emergence in a plage region, with a configuration similar to \citet{Martinez-Sykora+etal2017}. The horizontal resolution of their simulations was 14 km and 16 km, respectively, and the results of the simulations with and without non-equilibrium ionization were compared at about 12-14 minutes after the bifurcation. It was observed that initially cool bubbles and wakes of shocks remain over-ionized due to the imbalance between ionization and recombination scales. However, by the end of the time series, the ionization fraction begins to decrease as hydrogen and helium start to recombine. A longer time series may be needed to fully understand this process. Similar to the current work, \citet{Martinez-Sykora+etal2020} evaluated the scales at which ambipolar diffusion is active using Eq. ~\ref{eq:la}. They found values of $L_A$ up to $10^5$ m in the coldest and most rarefied areas in their LTE simulations. When non-equilibrium ionization is present, these large $L_A$ values extend over a broader range of temperatures and densities, corresponding to the upper chromosphere and transition region. To compare with their results, Figure \ref{fig:la_te_dens} shows the ambipolar scale $L_A$ as a function of density and temperature in all the models. We observe a behavior similar to \citet{Martinez-Sykora+etal2020}, except that our models cover a shorter height range and therefore fewer density values. Increasing the resolution amplifies the range of $T/\rho$ values corresponding to large $L_A$ (black and dark violet colors, compare panels 1-3). In the unipolar models (panels 4-6), the resolved points correspond to much higher temperatures, especially in the Bz200 case, as already seen from Figure \ref{fig:temp_height_eta}. Further work is needed to determine the extent to which these results will be affected by non-equilibrium ionization. Our models are run in 3D, where it would be extremely costly to include non-equilibrium ionization \citep[see, however][]{Martinez-Sykora+etal2023}. Nevertheless, while the amount of heating may indeed change, the main conclusion would still hold: ambipolar diffusion provides the largest scales among all possible dissipation mechanisms and therefore is the most important one in chromospheric layers.

Simulations of magnetic flux emergence in a plage region, with an average field strength in the photosphere of about 190 G and 14 km horizontal resolution by \citet{Martinez-Sykora+etal2017}, reveal higher average temperatures from the middle photosphere to the transition region, but lower temperatures in the corona in the models when ambipolar diffusion is acting. The cooler coronal temperatures were explained as a consequence of ambipolar diffusion dissipating free magnetic energy at the lower layers, making it less available at the coronal layers. The authors compared the Joule heating caused by the Ohmic-like hyperdiffusion term, the artificial diffusion needed to stabilize runs with ambipolar diffusion, and the proper ambipolar diffusion heating. Similar to the current work, \citet{Martinez-Sykora+etal2017} concluded that ambipolar and numerical heating act differently and in different areas, with ambipolar diffusion primarily heating the coolest areas of their domain. Though the ambipolar diffusion coefficient ($\eta_A$) in \citet{Martinez-Sykora+etal2017} was appreciably larger than the numerical equivalents in the chromosphere, this did not reflect in the Joule heating by ambipolar diffusion, which was only above the numerical one in some isolated locations. This goes in line with what was shown here in Figs.~\ref{fig:la-scales} and \ref{fig:la-image},  demonstrating that $Q_{\rm AMB}$ and $\eta_A$ distributions are different and that the $\eta_A$-base criterion is fulfilled in a significantly larger fraction of points compared to the $Q_{\rm AMB}$ criterion. Unfortunately, it was not specified in  \citet{Martinez-Sykora+etal2017} in which volume fraction the ambipolar effect is numerically resolved. The comparison of the numerical and ambipolar diffusion coefficients was also performed in \citet{MartinezSykora+etal2012}, showing $\eta_A$ above the numerical one in the cool areas of the chromosphere in simulations with the largest magnetization (90 G). In our work, we observe that locations with ambipolar heating above the numerical one generally coincide with locations where the ambipolar scale is resolved in the SSD simulations. However, this criterion is much less restrictive in the unipolar simulations. Therefore, as a general rule, both criteria need to be verified to conclude if the physical effect is resolved.

\begin{acknowledgements}
This work was supported by the Spanish Ministry of Science through the project PID2021-127487NB-I00 and by the European Research Council through the project ERC-2017-CoG771310-PI2FA. NV is funded by the European Union ERC AdG SUBSTELLAR grant agreement number 101054354. We thankfully acknowledge the technical expertise and assistance provided by the Spanish Supercomputing Network (Red Española de Supercomputación), as well as the computer resources used: LaPalma Supercomputer, located at the Instituto de Astrofísica de Canarias, and MareNostrum based in Barcelona/Spain.
\end{acknowledgements}

\providecommand{\noopsort}[1]{}\providecommand{\singleletter}[1]{#1}%

\end{document}